\documentclass[twocolumn,pra,showpacs,nofootinbib]{revtex4}%
\usepackage[pdftex,bookmarks]{hyperref}
\usepackage{amsfonts}
\usepackage{amsmath}
\usepackage{amssymb}
\usepackage{graphicx}%
\providecommand{\U}[1]{\protect\rule{.1in}{.1in}}
\begin{document}
\title{Quantum Repeater with Encoding}
\date{\today}
\author{Liang~Jiang$^{1}$}
\author{J.~M.~Taylor$^{2}$}
\author{Kae~Nemoto$^{3}$}
\author{W.~J.~Munro$^{3,4}$}
\author{Rodney~Van~Meter$^{3,5}$}
\author{M.~D.~Lukin$^{1}$}
\affiliation{$^{1}$Department of Physics, Harvard University, Cambridge, MA 02138, USA}
\affiliation{$^{2}$Department of Physics, Massachusetts Institute of Technology, Cambridge,
MA 02139, USA}
\affiliation{$^{3}$National Institute of Informatics, 2-1-2 Hitotsubashi, Chiyoda-ku, Tokyo
101-8430, Japan}
\affiliation{$^{4}$Hewlett-Packard Laboratories, Filton Road, Stoke Gifford, Bristol BS34
8QZ, UK}
\affiliation{$^{5}$Faculty of Environment and Information Studies, Keio University 5322
Endo, Fujisawa, Kanagawa 252-8520, Japans}

\begin{abstract}
We propose a new approach to implement quantum repeaters for long distance
quantum communication. Our protocol generates a backbone of encoded Bell pairs
and uses the procedure of classical error correction during simultaneous
entanglement connection. We illustrate that the repeater protocol with simple
Calderbank-Shor-Steane (CSS) encoding can significantly extend the
communication distance, while still maintaining a fast key generation rate.

\end{abstract}

\pacs{PACS number}
\maketitle


\section{Introduction}

\begin{figure}[t]
\begin{center}
\includegraphics[
width=8 cm
]{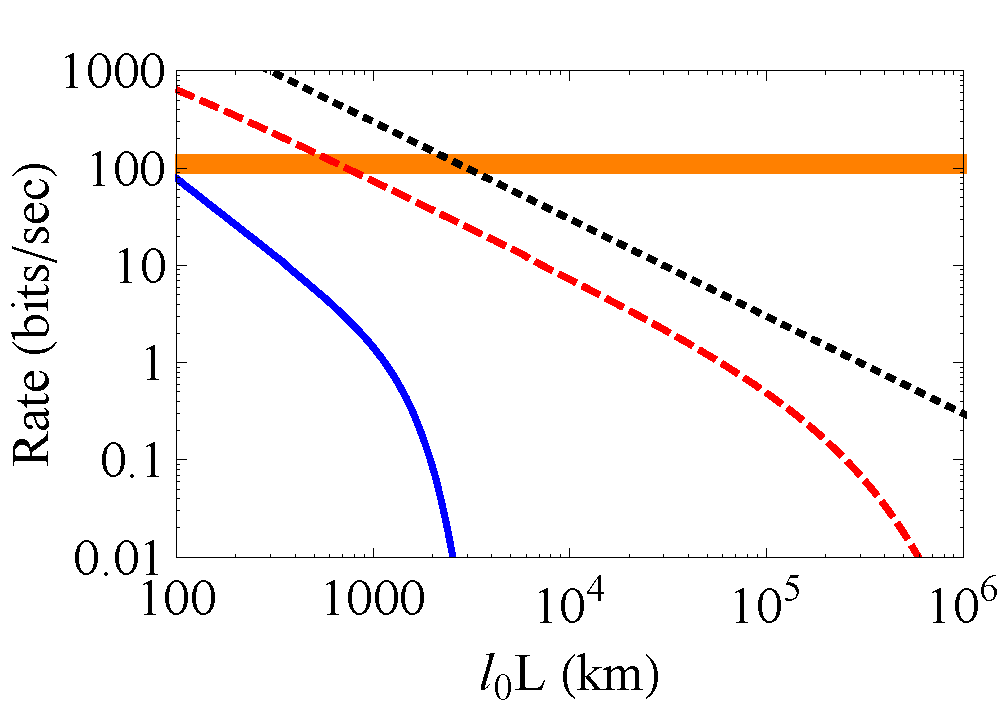}
\end{center}
\caption[fig:Comparison]{Comparison between the conventional and new repeater
protocols, in terms of the generation rate of Bell pairs or secret bit pairs
(i.e., the sustained bandwidth of the repeater channel). The memory coherence
time is assumed to be $t_{coh}=10$ sec, and the nearest neighbor spacing is
$l_{0}=10$ km. (i) The blue curve is the BDCZ protocol (with the maximum
number of qubits per station increasing logarithmically with distance, see
scheme C in Ref.~\cite{Dur99}). The sharp decrease in rate is attributed to
blinded connection and purification \cite{HartmannL07} when the memory error
becomes dominant (i.e., $\mathrm{time}\gtrsim0.01\tau_{coh}=0.1$ sec). (ii)
The red dashed curve is the parallel protocol (with the number of qubits per
station increasing at least linearly with distance, see scheme B in
Ref.~\cite{Dur99}). (iii) The black dotted reference curve is the inverse of
the classical communication time between the final stations. Since all
conventional repeater protocols rely on two-way classical communication, their
rates always stay below the reference curve unless parallel or multiplexed
\cite{VanMeter07c} repeater channels are used. (iv) The orange horizontal
thick line is our new repeater protocol with encoding (with the number of
qubits per station increasing logarithmically or poly-logarithmically with
distance). Since our protocol runs in the one-way communication mode, the rate
is independent of the communication distance, and can reach above the black
dashed curve. Our protocol is much more efficient over long distances than
conventional protocols.}%
\label{fig:Comparison}%
\end{figure}

Quantum key distribution generates a shared string of bits between two distant
locations (a key) whose security is ensured by quantum mechanics rather than
computational complexity \cite{Gisin02}. Recently, quantum communication over
150 km has been demonstrated \cite{Ursin07}, but the key generation rate
decreases exponentially with the distance due to the fiber attenuation.
Quantum repeaters can resolve the fiber attenuation problem, reducing the
exponential scaling to polynomial scaling by introducing repeater stations to
store intermediate quantum states \cite{Briegel98, Childress06b, vanLoock06}.
Dynamic programming-based search algorithm can optimize the key generation
rate and the final-state fidelity of the quantum repeaters \cite{JTKL07}.
Using additional local resources (i.e., more quantum bits per station), the
key generation rate can be further improved by multiplexing different
available pairs \cite{Collins07} and banding pairs according to their
fidelities \cite{VanMeter07c}. However, since all these protocols use
entanglement purification that requires two-way classical communication, the
time to purify pairs increases with the distance and all these protocols are
relatively slow. Thus, the finite coherence time of quantum memory ultimately
limits the communication distance \cite{HartmannL07}. As illustrated in
Fig.~\ref{fig:Comparison}, the estimated key generation rate sharply decreases
as soon as the memory error becomes dominant.

Here we propose a new, fast quantum repeater protocol in which the
communication distance is \emph{not} limited by the memory coherence time. Our
protocol encodes logical qubits with small CSS codes \cite{NC00}, applies
entanglement connection at the encoded level, and uses classical error
correction to boost the fidelity of entanglement connection. We eliminate the
time-consuming entanglement purification operation over long distances and
also avoid the resource-consuming procedure of quantum error correction. We
find that the new repeater protocol with small CSS codes can extend the
communication distance ($10^{3}\sim10^{6}$ km) and maintain an efficient key
generation rate (above $100$ bits/sec) using finite local resources
($30\sim150$ qubits/station) that scale logarithmically with distance.

In Sec.~\ref{sec:IdealizedRepeater}, we describe the idealized quantum
repeater protocol to overcome the fiber attenuation problem, emphasizing the
possibility of simultaneous entanglement connection and pointing out three
other major imperfections (entanglement infidelity, operational errors, and
memory errors) that still needs to be suppressed. In
Sec.~\ref{sec:QRwithRepetitionCode}, we consider an example of quantum
repeater with repetition code to suppress the bit-flip errors. In
Sec.~\ref{sec:QRwithCSS}, we provide the general protocol for quantum repeater
with CSS code that can suppress both bit-flip and dephasing errors. In
Sec.~\ref{sec:ErrorEstimate}, we compute the final fidelity achievable with
our protocol, which in principle can be arbitrarily close to unity using large
and efficient CSS code. In Sec.~\ref{sec:ExampleImplementations}, we calculate
the maximum number of connections depending on the code and the imperfections,
and we also estimate the key generation rate. In Sec.~\ref{sec:Discussion}, we
discuss potential improvements and other applications.

\section{Fast quantum communication with ideal
operations\label{sec:IdealizedRepeater}}

\begin{figure}[t]
\begin{center}
\includegraphics[
width=8.7 cm
]{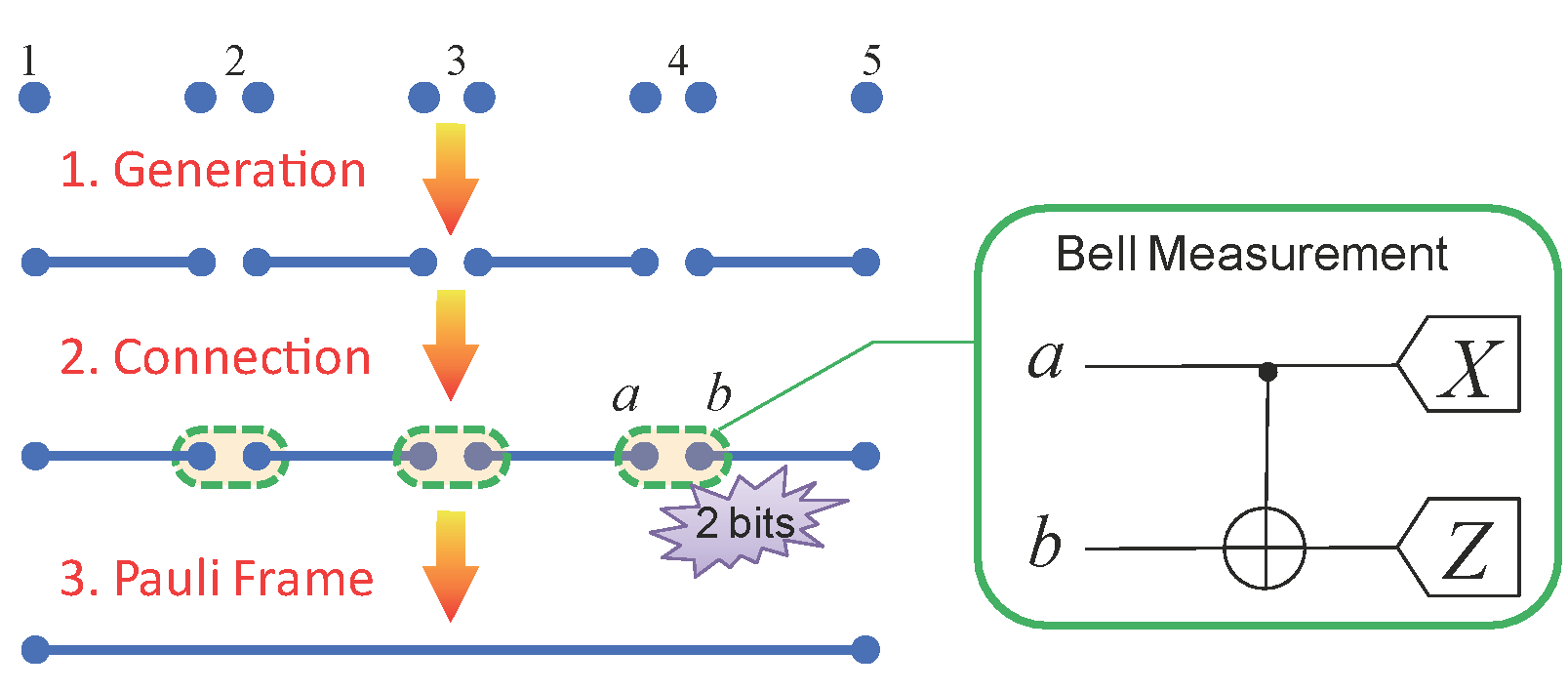}
\end{center}
\caption[fig:IdealQR]{Idealized quantum repeater. There are $L=5$ repeater
stations. Each intermediate station has two physical qubits. \textbf{Step 1.
(Generation)} Bell pairs between neighboring repeater stations are generated.
\textbf{Step} \textbf{2. (Connection)} The qubits at the intermediate stations
are measured in the Bell basis (see the inset). \textbf{Step 3. (Pauli Frame)}
The Pauli frame for qubits at the outermost stations is determined, based on
the outputs of intermediate Bell measurements. Finally, one remote Bell pair
between the outermost stations is created.}%
\label{fig:IdealQR}%
\end{figure}

We start by describing an idealized quantum repeater protocol, where fiber
attenuation is the only problem to be overcome. As illustrated in
Fig.~\ref{fig:IdealQR}, there are L repeater stations, and the separation
between the neighboring stations is of the order of the fiber attenuation
length. The Bell pairs $\left\vert \Phi^{+}\right\rangle =\frac{1}{\sqrt{2}%
}\left(  \left\vert 0\right\rangle \left\vert 0\right\rangle +\left\vert
1\right\rangle \left\vert 1\right\rangle \right)  $ between neighboring
repeater stations are independently generated and verified.
Then \emph{entanglement connection} (swapping) \cite{Bennett93, Zukowski93} is
applied to connect these Bell pairs into a long Bell pair. Each intermediate
repeater station measures the two local qubits in the Bell basis (called
\emph{Bell measurement}, see the inset of Fig.~\ref{fig:IdealQR}) and
announces $2$ classical bits of information, which uniquely specify the four
possible measurement outcomes and enables the determination of the \emph{Pauli
frame} for the remaining qubits (i.e., the choice of local Pauli operators
that adjust the long Bell pair to $\left\vert \Phi^{+}\right\rangle $
\cite{Knill05}). This is a deterministic process requiring local operation and
(one-way) classical communication.

The entanglement connection can be applied \emph{simultaneously} \footnote{We
use the rest frame of the repeater stations.}\ for all intermediate stations,
because the quantum circuit for Bell measurement does not depend on the Pauli
frame. It is the interpretation of the measurement outcome that depends on the
Pauli frame. Fortunately, we can wait until we collect all $2(L-2)$ announced
classical bits from intermediate stations, and decide the Pauli frame for the
final distant Bell pair. In addition, without compromising the security for
quantum key distribution, the two final (outermost) stations can measure their
qubits in random X and Z basis and announce their choices of the basis even
before receiving classical bits from intermediate stations. Half of the time,
they will find that they choose the same basis (in the Pauli frame) and obtain
strongly correlated measurement outcomes that can be used for secret keys
\cite{Ekert91}. Thanks to the \emph{simultaneous entanglement connection}, the
idealized quantum repeater can be very fast and the \emph{cycle time}
$\tau_{c}$ is just the total time for entanglement generation and connection
between neighboring repeater stations.

In practice, however, there are three major imperfections besides the fiber
attenuation. (1) The generated entangled state $\rho$ between neighboring
repeater stations is not the perfect Bell state $\left\vert \Phi
^{+}\right\rangle $, characterized by the entanglement fidelity
\begin{equation}
F_{0}=\left\langle \Phi^{+}\right\vert \rho\left\vert \Phi^{+}\right\rangle
\leq1.
\end{equation}
(2) The local operations for entanglement connection have errors
\cite{Briegel98, Childress06b, vanLoock06,JTKL07}. For example, the local
two-qubit unitary operation $U_{ij}$ would be%
\begin{equation}
U_{ij}\rho U_{ij}^{\dag}\rightarrow\left(  1-\beta\right)  U_{ij}\rho
U_{ij}^{\dag}+\frac{\beta}{4}\operatorname*{Tr}\nolimits_{ij}\left[
\rho\right]  \otimes I_{ij},
\end{equation}
where $\beta$ is the gate error probability, $\operatorname*{Tr}_{ij}\left[
\rho\right]  $ is the partial trace over the subsystem $i$ and $j$, and
$I_{ij}$ is the identity operator for the subsystem $i$ and $j$. The
projective measurement of state $\left\vert 0\right\rangle $ would be%
\begin{equation}
P_{0}=\left(  1-\delta\right)  \left\vert 0\right\rangle \left\langle
0\right\vert +\delta\left\vert 1\right\rangle \left\langle 1\right\vert ,
\end{equation}
where $\delta$ is the measurement error probability. (3) The quantum memory
decoheres with rate $\gamma$. We model the memory error probability for
storage time $\tau_{c}$ as $\mu=1-e^{-\gamma\tau_{c}}\approx\gamma\tau_{c}$.
The action of the memory error on the $i$th qubit would be%
\begin{equation}
\rho\rightarrow\left(  1-\mu\right)  \rho+\frac{\mu}{2}\operatorname*{Tr}%
\nolimits_{i}\left[  \rho\right]  \otimes I_{i},
\end{equation}
where $\operatorname*{Tr}_{i}\left[  \rho\right]  $ is the partial trace over
the subsystem $i$, and $I_{i}$ is the identity operator for the subsystem $i$.

\begin{figure*}[ptbh] \centering
\includegraphics[
width=15 cm
]{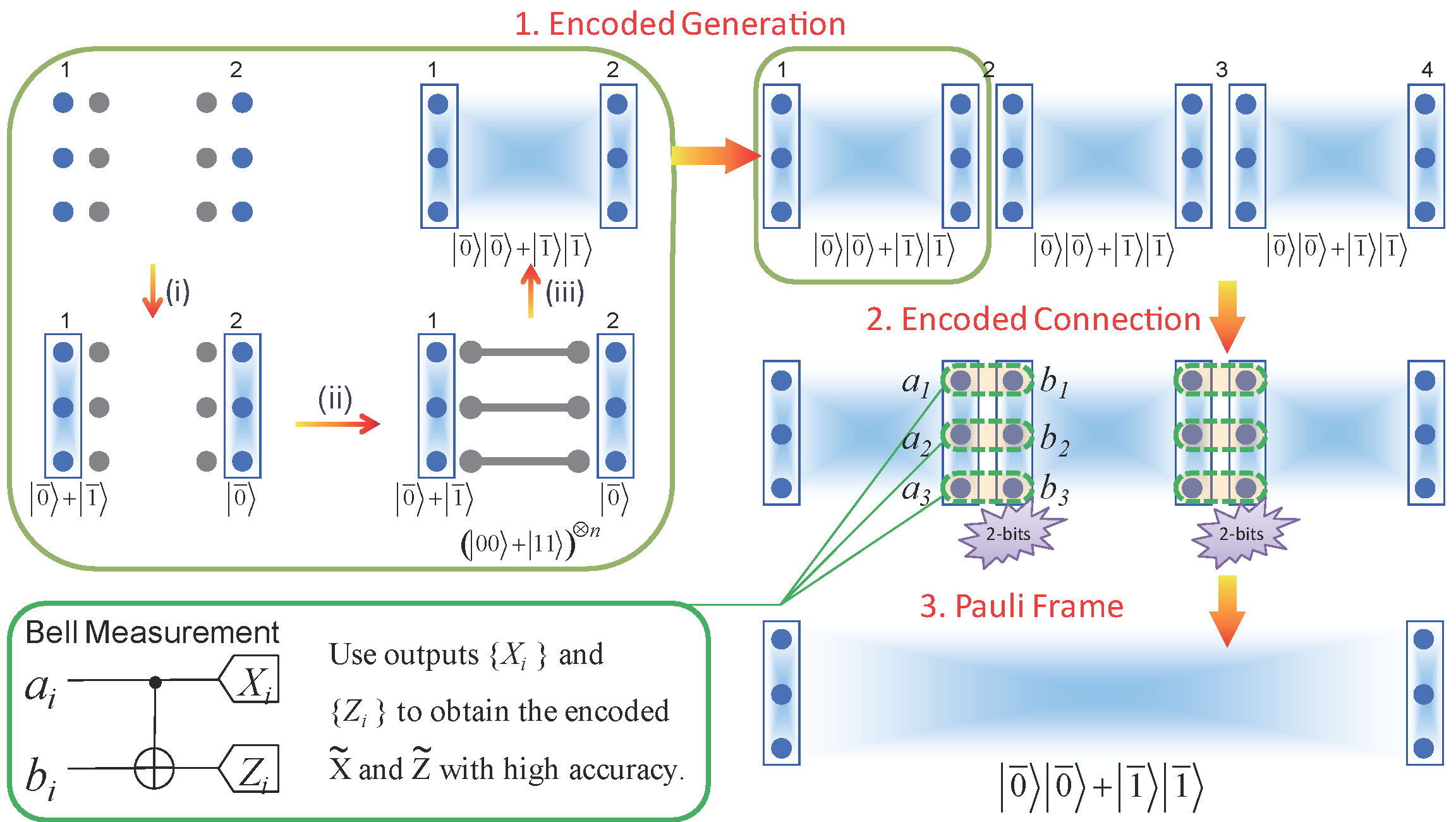}\caption[fig:EncodedQR]{Repeater protocol with
encoding.\textbf{ }Each repeater station has $2n$ memory qubits (blue dots)
and $O\left(  n\right)  $ auxiliary qubits (gray dots). Here $n=3$.
\textbf{Step 1. (Encoded Generation)} Between two neighboring stations
(upper-left panel): (i) memory qubits are fault-tolerantly prepared in the
encoded states $\left\vert \tilde{0}\right\rangle $ or $\left\vert \tilde
{+}\right\rangle =\frac{1}{\sqrt{2}}\left(  \left\vert \tilde{0}\right\rangle
+\left\vert \tilde{1}\right\rangle \right)  $, (ii) purified physical Bell
pairs are generated between auxiliary qubits (connected gray dots), (iii) an
encoded Bell pair $\left\vert \tilde{\Phi}^{+}\right\rangle _{AB}=\frac
{1}{\sqrt{2}}\left(  \left\vert \tilde{0}\right\rangle _{A}\left\vert
\tilde{0}\right\rangle _{B}+\left\vert \tilde{1}\right\rangle _{A}\left\vert
\tilde{1}\right\rangle _{B}\right)  $ between neighboring stations is created
using encoded CNOT\ gate (achieved by $n$ pairwise, teleportation-based
CNOT gates \cite{Gottesman99,Zhou00,JTSL07b}). \textbf{Step 2. (Encoded
Connection)} Encoded Bell measurements are simultaneously applied to all
intermediate repeater stations, via pairwise CNOT gates between
qubits $a_{i}$ and $b_{i}$ followed by
measurement of the physical qubits (the lower-left panel). Using classical
error correction, the outcomes for the encoded Bell measurement can be
obtained with a very small effective logical error probability $Q$ $\sim
q^{t+1}$ [Eq. (\ref{eq:LogicalErrorProb})]. The outcome is announced as $2$
classical bits (purple star) at each intermediate repeater station.
\textbf{Step 3. (Pauli Frame)} According to the outcomes of intermediate
encoded Bell measurements, the Pauli frame \cite{Knill05} can be determined for
qubits at the outermost stations. Finally, one encoded Bell pair between the
final (outermost) stations is created.}\label{fig:EncodedQR}%
\end{figure*}

In the following two sections, we will present the new repeater protocol. Our
new repeater protocol replaces the physical qubits (in Fig.~\ref{fig:IdealQR})
with encoded qubits (in Fig.~\ref{fig:EncodedQR}), generates the encoded Bell
pairs between neighboring stations, connects the encoded Bell pairs at
intermediate stations simultaneously, and determines the Pauli frame for the
encoded Bell pair shared by the final stations. In
Sec.~\ref{sec:QRwithRepetitionCode} we provide an illustrative example of
quantum repeater with $3$-qubit repetition code that can fix only bit-flip
errors. In Sec.~\ref{sec:QRwithCSS} we propose our new protocol with CSS codes
that can fix all imperfections listed above.

%

\section{Quantum repeater with repetition code\label{sec:QRwithRepetitionCode}%
}

To illustrate the idea, we first consider an example that uses the $3$-qubit
repetition code to encode one logical qubit%
\begin{equation}
\left\vert \tilde{0}\right\rangle =\left\vert 000\right\rangle \text{ \ and
\ }\left\vert \tilde{1}\right\rangle =\left\vert 111\right\rangle ,
\label{eq:RepetitionEncoding}%
\end{equation}
which can fix one bit-flip error. Although it cannot fix all the errors given
in Sec.~\ref{sec:IdealizedRepeater}, this example illustrates all other key
elements of the new repeater protocol and it can be easily generalized to the
CSS encodes that can fix all the errors as discussed in
Sec.~\ref{sec:QRwithCSS}.

First, we generate the encoded Bell pair $\left\vert \tilde{\Phi}%
^{+}\right\rangle _{12}=\frac{1}{\sqrt{2}}\left(  \left\vert \tilde
{0}\right\rangle _{1}\left\vert \tilde{0}\right\rangle _{2}+\left\vert
\tilde{1}\right\rangle _{1}\left\vert \tilde{1}\right\rangle _{2}\right)  $
between neighboring stations $1$ and $2$, as illustrated in the upper-left
panel of Fig.~\ref{fig:EncodedQR}. We need at least six qubits from each
station: three for memory qubits (blue dots) and three for ancillary qubits
(gray dots). There are three steps: (i) We locally prepare the encoded state
$\frac{1}{\sqrt{2}}\left(  \left\vert \tilde{0}\right\rangle _{1}+\left\vert
\tilde{1}\right\rangle _{1}\right)  $ and $\left\vert \tilde{0}\right\rangle
_{2}$ and store them in the memory qubits (in blue squared boxes); (ii) we
generate three copies of the physical Bell pairs $\left(  \frac{\left\vert
0\right\rangle _{1}\left\vert 0\right\rangle _{2}+\left\vert 1\right\rangle
_{1}\left\vert 1\right\rangle _{2}}{\sqrt{2}}\right)  ^{\otimes3}$ between
ancillar qubits (gray lines); (iii) we use the entanglement resources of $3$
physical Bell pairs to implement $3$ \emph{teleportation-based CNOT gates}
\cite{Gottesman99,Zhou00,JTSL07b}, applied \emph{transversally} between the
memory qubits storing the encoded states $\frac{1}{\sqrt{2}}\left(  \left\vert
\tilde{0}\right\rangle _{1}+\left\vert \tilde{1}\right\rangle _{1}\right)  $
and $\left\vert \tilde{0}\right\rangle _{2}$:%
\begin{align}
&  \frac{1}{\sqrt{2}}\left(  \left\vert 000\right\rangle _{1}+\left\vert
111\right\rangle _{1}\right)  \otimes\left\vert 000\right\rangle _{2}\\
&  \rightarrow\frac{1}{\sqrt{2}}\left(  \left\vert 000\right\rangle
_{1}\left\vert 000\right\rangle _{2}+\left\vert 111\right\rangle
_{1}\left\vert 111\right\rangle _{2}\right)  ,
\end{align}
which gives us exactly the desired encoded Bell pair $\left\vert \tilde{\Phi
}^{+}\right\rangle _{12}$. Similarly, we can generate encoded Bell pairs
$\left\vert \tilde{\Phi}^{+}\right\rangle _{j,j+1}$ between neighboring
stations $j$ and $j+1$, for $j=2,\cdots,L-1$.

Then we connect the encoded Bell pairs $\left\vert \tilde{\Phi}^{+}%
\right\rangle _{12}$ and $\left\vert \tilde{\Phi}^{+}\right\rangle _{23}$ to
obtain the longer encoded Bell pair $\left\vert \tilde{\Phi}^{+}\right\rangle
_{13}$. The idea is to perform the encoded Bell measurement over the two
encoding blocks at station $2$. We use $2a$ and $2b$ to refer to the left and
the right encoding blocks at station $2$, respectively. As shown in
Fig.~\ref{fig:EncodedQR} (see step 2 and the lower-left panel), we apply three
pairwise CNOT gates between the two encoding blocks $\left\{  a_{i}\right\}  $
and $\left\{  b_{i}\right\}  $ at station $2$. To see the possible outcomes of
this procedure, we rewrite the initial state in terms of Bell states between
stations $1$ and $3$,%
\begin{align*}
&  \left\vert \tilde{\Phi}^{+}\right\rangle _{1,2a}\otimes\left\vert
\tilde{\Phi}^{+}\right\rangle _{2b,3}\\
=  &  \frac{1}{2}\left(
\begin{array}
[c]{c}%
\left\vert \tilde{\Phi}^{+}\right\rangle _{13}\otimes\left\vert \tilde{\Phi
}^{+}\right\rangle _{2a,2b}+\left\vert \tilde{\Phi}^{-}\right\rangle
_{13}\otimes\left\vert \tilde{\Phi}^{-}\right\rangle _{2a,2b}\\
+\left\vert \tilde{\Psi}^{+}\right\rangle _{13}\otimes\left\vert \tilde{\Psi
}^{+}\right\rangle _{2a,2b}+\left\vert \tilde{\Psi}^{-}\right\rangle
_{13}\otimes\left\vert \tilde{\Psi}^{-}\right\rangle _{2a,2b}%
\end{array}
\right) \\
\rightarrow &  \frac{1}{2}\left(
\begin{array}
[c]{c}%
\left\vert \tilde{\Phi}^{+}\right\rangle _{13}\otimes\left\vert \tilde
{+}\right\rangle _{2a}\left\vert \tilde{0}\right\rangle _{2b}+\left\vert
\tilde{\Phi}^{-}\right\rangle _{13}\otimes\left\vert \tilde{-}\right\rangle
_{2a}\left\vert \tilde{0}\right\rangle _{2b}\\
+\left\vert \tilde{\Psi}^{+}\right\rangle _{13}\otimes\left\vert \tilde
{-}\right\rangle _{2a}\left\vert \tilde{0}\right\rangle _{2b}+\left\vert
\tilde{\Psi}^{-}\right\rangle _{13}\otimes\left\vert \tilde{-}\right\rangle
_{2a}\left\vert \tilde{1}\right\rangle _{2b}%
\end{array}
\right)  ,
\end{align*}
\newline where $\left\vert \tilde{\Phi}^{\pm}\right\rangle _{13}=\frac
{1}{\sqrt{2}}\left(  \left\vert \tilde{0}\right\rangle _{1}\left\vert
\tilde{0}\right\rangle _{3}\pm\left\vert \tilde{1}\right\rangle _{1}\left\vert
\tilde{1}\right\rangle _{3}\right)  $, $\left\vert \tilde{\Psi}^{\pm
}\right\rangle _{13}=\frac{1}{\sqrt{2}}\left(  \left\vert \tilde
{0}\right\rangle _{1}\left\vert \tilde{1}\right\rangle _{2}\pm\left\vert
\tilde{1}\right\rangle _{1}\left\vert \tilde{0}\right\rangle _{2}\right)  $,
$\left\vert \tilde{\pm}\right\rangle _{2a}=\frac{1}{\sqrt{2}}\left(
\left\vert \tilde{0}\right\rangle _{2a}\pm\left\vert \tilde{1}\right\rangle
_{2a}\right)  $. To complete the encoded Bell measurement, we projectively
measure the logical qubits of these two encoding blocks as follows: (1) The
logical qubit for $2a$ should be measured in the \{$\left\vert \tilde{\pm
}\right\rangle $\} basis, which can be achieved by measuring the physical
qubits $\left\{  a_{i}\right\}  $ in the \{$\left\vert \pm\right\rangle $\}
basis. Since $\left\vert \tilde{+}\right\rangle =\frac{1}{2}\left(  \left\vert
+++\right\rangle +\left\vert +--\right\rangle +\left\vert -+-\right\rangle
+\left\vert --+\right\rangle \right)  $ and $\left\vert \tilde{-}\right\rangle
=\frac{1}{2}\left(  \left\vert ---\right\rangle +\left\vert -++\right\rangle
+\left\vert +-+\right\rangle +\left\vert ++-\right\rangle \right)  $, there
will be an odd number of $\left\vert +\right\rangle $ outputs if the encoded
qubit is in state $\left\vert \tilde{+}\right\rangle $, and an even number of
$\left\vert +\right\rangle $ outputs if the encoded qubit is in state
$\left\vert \tilde{-}\right\rangle $. (2) The logical qubit for $2b$ should be
measured in the \{$\left\vert \tilde{0}\right\rangle ,\left\vert \tilde
{1}\right\rangle $\} basis, which can be achieved by measuring the physical
qubits $\left\{  b_{i}\right\}  $ in the \{$\left\vert 0\right\rangle
,\left\vert 1\right\rangle $\} basis. There should be three $\left\vert
0\right\rangle $ outputs for state $\left\vert \tilde{0}\right\rangle $, and
three $\left\vert 1\right\rangle $ outputs for state $\left\vert \tilde
{1}\right\rangle $. The pairwise CNOT gates and projective measurement of
physical qubits are summarized in the lower-left panel of
Fig.~\ref{fig:EncodedQR}.

We now show the suppression of bit-flip errors due to the repetition code
[Eq.~(\ref{eq:RepetitionEncoding})]. If one of the physical qubits in $2b$ is
bit-flipped, the measurement outcomes for $2b$ will contain two correct
outputs and one erroneous output. Choosing the majority output, we can
identify and correct the erroneous output, and still obtain the logical bit
encoded in $2b$ correctly. We emphasize that only classical error correction
is used. If there is one physical qubit in $2a$ that suffers from a bit-flip
error, this error will not affect the outputs for $2a$, as bit-flip errors
commute with the operators to be measured; this error may affect one
corresponding physical qubit in $2b$, which can be identified and corrected
using the majority. Therefore, we can obtain the logical outcomes for both
$2a$ and $2b$, and the suppressed effective logical error probability can be%
\begin{equation}
Q=\left(
\begin{array}
[c]{c}%
3\\
2
\end{array}
\right)  q_{b}^{2}+\left(
\begin{array}
[c]{c}%
3\\
3
\end{array}
\right)  q_{b}^{3}\approx6q_{b}^{2}\ll q_{b},
\end{equation}
where $q_{b}\leq4\beta+2\delta+\mu$ is the effective error probability for
each $b_{i}$ to give the wrong output (Appendix.~\ref{app:EffectiveErrorProb}).

To complete the entanglement connection, station $2$ announces the outcomes
for its two logical qubits from the encoded Bell measurement, which contains
\emph{two} classical bits of information and determines the Pauli frame for
the encoded Bell pair shared between stations $1$ and $3$. Note that the
detailed outputs of physical qubits are only important to obtain the logical
outcomes, but not needed for communication among stations. Similarly, we can
perform entanglement connection for all the intermediate stations.
Furthermore, these entanglement connections can still be applied
simultaneously for all intermediate striations, as described in
Sec.~\ref{sec:IdealizedRepeater}. The final stations share the encoded Bell
pair $\left\vert \tilde{\Phi}^{+}\right\rangle _{1L}=\frac{1}{\sqrt{2}}\left(
\left\vert \tilde{0}\right\rangle _{1}\left\vert \tilde{0}\right\rangle
_{L}+\left\vert \tilde{1}\right\rangle _{1}\left\vert \tilde{1}\right\rangle
_{L}\right)  $, whose Pauli frame is determined by the $2(L-2)$ announced
classical bits from all intermediate stations.

\section{Quantum repeater with CSS code\label{sec:QRwithCSS}}

In this section, we will generalize the repeater protocol from the $3$-qubit
repetition code to any $[[n,k,2t+1]]$ \emph{CSS code} \cite{NC00}, which
encodes $k$ logical qubits with $n$ physical qubits and fixes up to $t$
(bit-flip and dephasing) errors. For simplicity, we will focus on the CSS
codes with $k=1$, which includes the well studied $\left[  \left[
5,1,3\right]  \right]  $, $\left[  \left[  7,1,3\right]  \right]  $ (Steane),
and $\left[  \left[  9,1,3\right]  \right]  $ (Shor) codes. Extension of the
protocol to $k>1$ is straightforward. The CSS code can be regarded as a
combination of two classical error correcting codes $C^{X}$ and $C^{Z}$, which
fix dephasing errors and bit-blip errors, respectively. The error syndromes
for the code $C^{X}$ (or $C^{Z}$) can be obtained if we have the outputs for
the $n$ physical qubits measured in the $X$ (or $Z$) basis.

The relevant properties of the CSS codes are summarized as follows: (1) The
measurement of logical operator $\tilde{X}$ (or $\tilde{Z}$) can be obtained
from projective measurement of physical qubits in the $X$ (or $Z$) basis. (2)
The outputs from measurements of physical qubit in the $X$ (or $Z$) basis
should comply with the rules of the classical error correcting code $C^{X}$
(or $C^{Z}$), which can fix up to $t^{X}$ (or $t^{Z}$) errors in the $n$
output bits. (For example, the $3$-qubit repetition cod can fix up to
$t^{Z}=1$ bit-flip error as discussed in Sec.~\ref{sec:QRwithRepetitionCode}.)
Suppose each output bit has an (uncorrelated) effective error probability
$q\sim\beta+\delta+\mu$, after fixing up to $t$ errors the remaining error
probability for the logical outcome is $O\left(  q^{t+1}\right)  $, assuming
$t=\min\left\{  t^{X},t^{Z}\right\}  $. (3) The encoded CNOT gate can be
implemented by $n$ pairwise CNOT gates between two encoding blocks
\cite{NC00}. Such pairwise CNOT\ gates do not propagate errors within each
encoding block, and it can be be used for preparation of encoded Bell pairs.

We find that each repeater station needs approximately $6n$ physical qubits
(see Appendix~\ref{app:FT initialization} for details), including $2n$ memory
qubits to store the two encoded qubits that are entangled with the neighboring
stations, and approximately $4n$ ancillary qubits for the fault-tolerant
preparation of the encoded qubits and generation of non-local encoded Bell
pairs between neighboring repeater stations. There are three steps for each
cycle of the new repeater protocol:

1. Generate encoded Bell pairs between two neighboring stations (see the
upper-left panel of Fig.~\ref{fig:EncodedQR}): (i) We initialize the memory
qubits in logical states $\left\vert \tilde{0}\right\rangle $ and $\frac
{1}{\sqrt{2}}\left(  \left\vert \tilde{0}\right\rangle +\left\vert \tilde
{1}\right\rangle \right)  $ at each station \emph{fault-tolerantly} (with
errors effectively uncorrelated among physical qubits from the same encoding
block) \footnote{We can achieve fault-tolerant preparation of the logical
state $\left\vert \tilde{0}\right\rangle $ (or $\left\vert \tilde
{+}\right\rangle $) by two approaches. One approach uses several copies of the
logical states to distill a purified logical state with negligible
contribution from initial correlated errors \cite{Steane99}. Alternatively, we
may start with $\left\vert 0\right\rangle ^{\otimes n}$ (or $\left\vert
+\right\rangle ^{\otimes n}$), projectively measure the x-stabilizers using
fault-tolerant circuit, and update the stabilizers during entanglement
connection. See Appendix~\ref{app:FT initialization} for details.}. (ii) We
use entanglement purification to obtain purified Bell pairs between two
neighboring stations \cite{Dur99}. Each purified Bell pair can be immediately
used for a \emph{teleportation-based CNOT gate}
\cite{Gottesman99,Zhou00,JTSL07b}. (iii) According to the property (3) of the
CSS code, we need $n$ teleportation-based CNOT gates to implement the encoded
CNOT gate and obtain the encoded Bell pair $\left\vert \tilde{\Phi}%
^{+}\right\rangle _{j,j+1}=\frac{1}{\sqrt{2}}\left(  \left\vert \tilde
{0}\right\rangle _{j}\left\vert \tilde{0}\right\rangle _{j+1}+\left\vert
\tilde{1}\right\rangle _{j}\left\vert \tilde{1}\right\rangle _{j+1}\right)  $
between neighboring stations $j$ and $j+1$.

2. Connect the encoded Bell pairs, performing encoded Bell measurement at all
intermediate stations simultaneously (see step 2 of Fig.~\ref{fig:EncodedQR}).
At each intermediate station, we first apply the pairwise CNOT gates between
qubits $a_{i}$ and $b_{i}$, with $a_{i}$ from the control block and $b_{i}$
from the target block, for $i=1,\cdots,n$ (as shown in the lower-left panel in
Fig.~\ref{fig:EncodedQR}). Then we projectively measure the physical qubits in
the $X$ basis for $a_{i}$ and in the $Z$ basis for $b_{i}$. According to the
property (2) of the CSS code, we can use the classical error correcting code
$C^{X}$ (or $C^{Z}$) to fix up to $t$ errors in $\left\{  a_{i}\right\}  $ (or
$\left\{  b_{i}\right\}  $), leaving only $O\left(  q^{t+1}\right)  $ for the
logical error probability. Thus, the outcomes for the encoded $\tilde{X}$ and
$\tilde{Z}$ operators of the encoded Bell measurement can be obtained with
high accuracy of $O\left(  q^{t+1}\right)  $. Similar to the idealized
repeater, each intermediate repeater station announces $2$ classical bits of
information of the encoded Bell measurement.

3. According to the $2(L-2)$ announced classical bits from all intermediate
repeater stations, choose the Pauli frame at the final repeater stations for
the shared encoded Bell pair.

\section{Error estimate\label{sec:ErrorEstimate}}

In order to calibrate the encoded Bell pair obtained from the new repeater
protocol, we need to generalize the definition of entanglement fidelity,
because the encoding enables us to correct small errors that deviate from the
logical subspace. We define the \emph{entanglement fidelity} as
\begin{equation}
F=\left\langle \tilde{\Phi}^{+}\right\vert \mathcal{R}\left[  \rho
_{\mathrm{fina,Bell}}\right]  \left\vert \tilde{\Phi}^{+}\right\rangle ,
\end{equation}
where $\mathcal{R}$ represents the (ideal) recovery operation with quantum
error correction \cite{Shor00}. The entanglement fidelity $F$ can calibrate
the security for the protocol and bound the maximum information leaked from
the final stations \footnote{According to Ref. \cite{Lo99}, if the two
final stations share a Bell pair with fidelity $F=1-2^{-s}$, then Eve's mutual
information with the key is at most $2^{-c}+2^{O\left(  -2s\right)  }$ where
$c=s-\log_{2}\left(  2+s+1/\ln2\right)  $.}. $F$ can also be practically
obtained from the correlation measurement between the final repeater station
(Appendix~\ref{app:Fidelity and Correlation}).

We emphasize that the property of fault-tolerance can be maintained throughout
the entire repeater protocol (fault-tolerant initialization, transverse CNOT
gate, and encoded qubit measurement), so the errors for individual physical
qubits are effectively uncorrelated. With some calculation (see
Appendix.~\ref{app:EffectiveErrorProb}), we estimate that the effective error
probability (per physical qubit) is
\begin{equation}
q=4\beta+2\delta+\mu,
\end{equation}
Note that $q$ does not explicitly depend on $F_{0}$, because for level-$m$
purified Bell pairs (see Appendix~\ref{app:TimeOverhead}) the operational
errors ($\beta$ and $\delta$) dominate the super-exponentially suppressed
infidelity approximately $\left(  1-F_{0}\right)  ^{2^{m/2}}$. Then the
\emph{effective logical error probability} for each encoding block (caused by
more than $t$ errors from the encoded block) is
\begin{equation}
Q=\sum_{j=t+1}^{n}\left(
\begin{array}
[c]{c}%
n\\
j
\end{array}
\right)  q^{j}\left(  1-q\right)  ^{n-j}\approx\left(
\begin{array}
[c]{c}%
n\\
t+1
\end{array}
\right)  q^{t+1},\label{eq:LogicalErrorProb}%
\end{equation}
where the approximation requires small $q$. Since any logical error from the
repeater stations will affect the final encoded Bell pair, the entanglement
fidelity is%
\begin{equation}
F=(1-Q)^{2L},\label{eq:Fidelity}%
\end{equation}
with infidelity $1-F\approx2LQ$ for small $Q$.


\begin{figure}[t]
\begin{center}
\includegraphics[
width=8.5cm
]{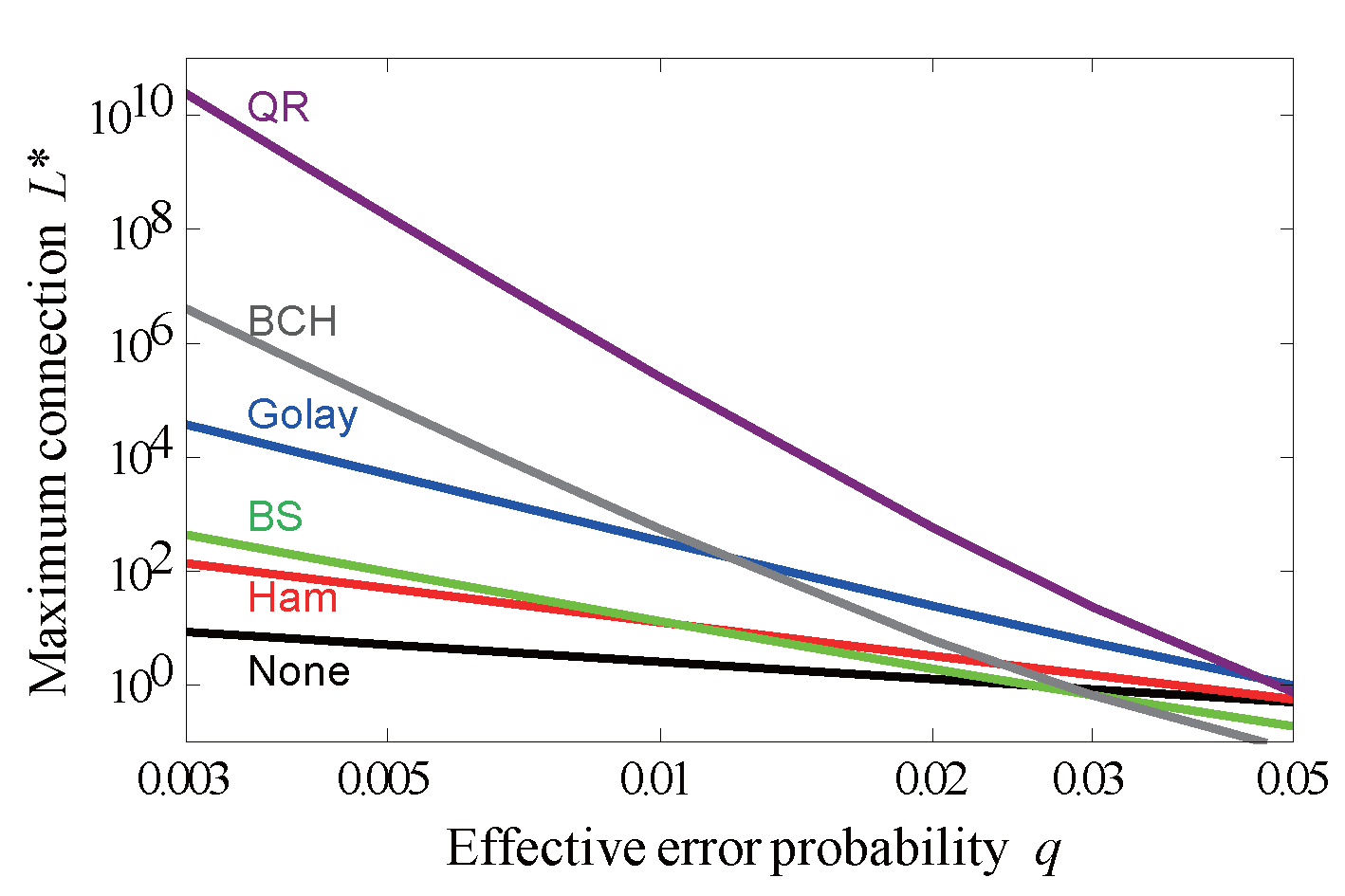}
\end{center}
\caption[fig:Example]{From Eqs.~(\ref{eq:LogicalErrorProb},\ref{eq:Lstar}),
the maximum number of connections $L^{\ast}$ is estimated as a function of the
effective error probability $q$, assuming $F^{\ast}=0.95$, for various CSS
codes \cite{Steane03} listed in Table~\ref{tab:1}. For $q<0.03$, $L^{\ast}$
scales as $1/q^{t+1}$.}%
\label{fig:Example}%
\end{figure}

For large codes, we may evaluate Eq.~(\ref{eq:LogicalErrorProb}) under the
assumptions $n\gg t\gg1$. Approximating the combinatorial function in this
limit yields $Q\approx\frac{1}{\sqrt{2\pi t}}\left(  \frac{e^{1+1/2n}nq}%
{t+1}\right)  ^{t+1}$, which indicates that for large codes with $n\propto t$,
$Q$ can be arbitrarily small when $q\lesssim q_{c}\approx\lim_{n,t\rightarrow
\infty}\frac{t+1}{e^{1+1/2n}n}$. Numerically, we can evaluate the complete sum
in Eq.~(\ref{eq:LogicalErrorProb}) and we find $q_{c}\approx5\%$, which
corresponds to $\sim1\%$ per-gate error rates. In addition, CSS codes with
$n\lesssim19t$ exist for arbitrarily large $t$ (according to the
Gilbert-Varsharov bound, see Eq.~(30) in Ref.~\cite{Canderbank96}). Therefore,
our new repeater protocol with encoding provides a scalable approach for long
distance quantum communication.

\section{Example implementations\label{sec:ExampleImplementations}}

We now consider the implementation of the new repeater protocol. Given the
effective error probability $q$ and the \emph{target fidelity} $F^{\ast}$, we
can use Eq.~(\ref{eq:Fidelity}) to calculate the \emph{maximum number of
connections}%

\begin{equation}
L^{\ast}=\frac{\ln F^{\ast}}{\ln\left(  1-Q\right)  }. \label{eq:Lstar}%
\end{equation}
This provides a unitless distance scale over which Bell pairs with fidelity
$F^{\ast}$ can be created. According to Eqs.~(\ref{eq:LogicalErrorProb}%
,\ref{eq:Lstar}), we can estimate $L^{\ast}$ as a function of $q$, which is
plotted in Fig.~\ref{fig:Example} assuming $F^{\ast}=0.95$ for various CSS
codes. Since $L^{\ast}$ scales as $q^{-\left(  t+1\right)  }$ for
$q\lesssim3\%$, we can significantly increase $L^{\ast}$ by considering
efficient quantum codes with large $t$. For example, given $q=0.3\%$, we
estimate the maximum number of connections $L^{\ast}\approx9$, $1.4\times
10^{2}$, and $3.7\times10^{4}$ for cases of no encoding, $\left[  \left[
7,1,3\right]  \right]  $ Hamming code, and $\left[  \left[  23,1,7\right]
\right]  $ Golay code, respectively. If we choose the nearest neighbor spacing
to be $l_{0}=10$ km (about half the fiber attenuation length), the
corresponding maximum distances will be $90$ km, $1.4\times10^{3}$ km, and
$3.7\times10^{5}$ km. The new protocol can easily reach and go beyond
intercontinental distances. In Table~\ref{tab:1}, we summarize the local
resources and maximum communication distance for the new protocol with
different encoding.%

\begin{table}[tbp] \centering
\begin{tabular}
[c]{ccll}\hline\hline
\textbf{Name} &
\begin{tabular}
[c]{c}%
\textbf{Code}\\
$\left[  \left[  n,k,2t+1\right]  \right]  $%
\end{tabular}
&
\begin{tabular}
[c]{c}%
\textbf{Resources}\\
{\small (qubits/station)}%
\end{tabular}
&
\begin{tabular}
[c]{c}%
\textbf{Distance}\\
{\small (km)}%
\end{tabular}
\\\hline
No encoding & -- & \multicolumn{1}{c}{$4$} & \multicolumn{1}{c}{$180$}\\\hline
Repetition-3 & $\left[  3,1,3\right]  $ & \multicolumn{1}{c}{$18$} &
\multicolumn{1}{c}{$1.0\times10^{4}$}\\
Repetition-5 & $\left[  5,1,5\right]  $ & \multicolumn{1}{c}{$30$} &
\multicolumn{1}{c}{$1.0\times10^{6}$}\\\hline
Hamming & $\left[  \left[  7,1,3\right]  \right]  $ & \multicolumn{1}{c}{$42$}
& \multicolumn{1}{c}{$1.4\times10^{3}$}\\
Bacon-Shor & $\left[  \left[  25,1,5\right]  \right]  $ &
\multicolumn{1}{c}{$150$} & \multicolumn{1}{c}{$4.3\times10^{3}$}\\
Golay & $\left[  \left[  23,1,7\right]  \right]  $ & \multicolumn{1}{c}{$138$}
& \multicolumn{1}{c}{$3.7\times10^{5}$}\\
BCH & $\left[  \left[  127,29,15\right]  \right]  $ & \multicolumn{1}{c}{--} &
\multicolumn{1}{c}{$4.0\times10^{7}$}\\
QR & $\left[  \left[  103,1,19\right]  \right]  $ & \multicolumn{1}{c}{--} &
\multicolumn{1}{c}{$2.4\times10^{11}$}\\\hline\hline
\end{tabular}
\newline%
\caption{Local resources, and maximum communication distance for the new repeater
protocol. In the case of no encoding, each station has $2$ qubits for entanglement connection,
and $2$ additional qubits for entanglement purification to obtain high-fidelity purified Bell pairs.
For repetition codes (with single square bracket), only
one type of errors (bit-flip or dephasing) can be suppressed.
For other CSS codes (with double
square brackets), both bit-flip and dephasing errors can be suppressed.
The local resources are estimated to
be $6n$ qubits for each station (Appendix~\ref{app:FT initialization}).
The distance is estimated from Eqs.~(\ref{eq:LogicalErrorProb},\ref{eq:Lstar}),
assuming parameters $q=0.3\%$, $F^{\ast}=0.95$ and $l_{0}=10$ km.}\label{tab:1}%
\end{table}%

Besides maximum distances, we also estimate the key generation rate, which is
the inverse of the cycle time for the new protocol. For fast local operations
(systems such as ion traps \cite{Leibfried04, Riebe04} and NV\ centers
\cite{Jelezko04,Dutt07} can achieve almost MHz rate for local operations), the
cycle time is dominated by creating purified Bell pairs between neighboring
stations%
\begin{equation}
\tau_{c}\approx\kappa\frac{l_{0}}{v}\frac{e^{l_{0}/l_{att}}}{\eta^{2}}.
\end{equation}
We find that $\tau_{c}\approx0.9\kappa$ ms, given the parameters of $l_{0}=10$
km, the fiber attenuation length $l_{att}\approx20$ km, the signal propagation
speed $v\approx2\times10^{5}$ km/s, and the overall efficiency for collecting
and detecting single photon $\eta\approx0.3$. The dimensionless prefactor
$\kappa$ is the time overhead to ensure that $n$ purified Bell pairs are
obtained between neighboring stations. Since there are approximately $4n$
ancillary qubits for entanglement generation at each station, the rate to
generate unpurified Bell pairs also increases with $n$. Thus, $\kappa$ is not
sensitive to the choice of $n$. As detailed in Appendix~\ref{app:TimeOverhead}%
, we estimate $\kappa\approx8$ for $\beta=\delta=10^{-3}$ and $F_{0}=0.95$
with depolarizing error, and the purified pair has fidelity $0.9984$ after
three levels of purification. Therefore, for the parameters considered here,
approximately $6n$ qubits at each station can achieve $\tau_{c}\approx7$ ms,
which is sufficient for quantum key generation rate of $100$ bits/sec over
long distances.

\section{Discussion\label{sec:Discussion}}

Our new repeater protocol is significantly faster than the standard repeater
protocols over long distances \cite{Briegel98, Childress06b, vanLoock06,
JTKL07}, because the time-consuming procedure of entanglement purification of
distant Bell pairs is now replaced by local encoding with simple CSS code and
classical error correction. The new protocol runs in the one-way communication
mode, so the key generation rate is independent of the communication distance,
and only limited by the cycle time for encoded Bell pair generation and
entanglement connection. The key generation rate can be further improved by
having higher efficiency $\eta$, improved fidelity $F_{0}$, smaller separation
between stations $l_{0}$, and more qubits per repeater station. In addition,
the rate can also be increased by using CSS codes with $k>1$ (e.g., the
$\left[  \left[  127,29,15\right]  \right]  $ BCH code mentioned in
Fig.~\ref{fig:Example}), along with a small modification to the protocol that
each intermediate station sends $2k$ classical bits associated with $k$ Bell measurements.

Asymptotically, CSS codes with $n\lesssim19t$ exist for arbitrarily large $t$
[obtained from the Gilbert-Varsharov bound, see Eq.~(30) in Ref.
\cite{Canderbank96}]. Thus, the effective logical error probability $Q$
[Eq.~(\ref{eq:LogicalErrorProb})] can be arbitrarily small for $q\lesssim5\%$,
and $n\propto t\sim\ln L$ is a small number increasing only logarithmically
with $L$.\textrm{ }In practice, however, it is still challenging to initialize
large CSS\ encoding block fault-tolerantly with imperfect local operations. To
avoid complicated initialization, we may construct larger CSS\ codes by
concatenating smaller codes with $r$ nesting levels, and the code size scales
polynomially with the code distance, $n\propto t^{r}\sim\left(  \ln L\right)
^{r}$. Alternatively, we may consider the Bacon-Shor code \cite{Bacon06}; the
encoding block scales quadratically with the code distance $n=\left(
2t+1\right)  ^{2}\sim\ln^{2}L$, and the initialization can be reduced to the
preparation of $\left(  2t+1\right)  $-qubit GHZ\ states.

If the imperfections are dominated by the dephasing errors, we may use the
$\left[  2t+1,1,2t+1\right]  $ repetition code [e.g., use the $\left(
2t+1\right)  $-qubit GHZ\ states $\left\vert +\cdots+\right\rangle
\pm\left\vert -\cdots-\right\rangle $ with $\left\vert \pm\right\rangle
=\frac{1}{\sqrt{2}}\left(  \left\vert 0\right\rangle \pm\left\vert
1\right\rangle \right)  $ to encode one logical qubit]. The repetition code
has the advantage of small encoding block and efficient initialization (see
Table~\ref{tab:1}). For example, given $q=0.3\%$ and $F^{\ast}=0.95$, we
estimate $L^{\ast}\approx1.0\times10^{3}$ and $1.0\times10^{5}$ for $3$-qubit
and $5$-qubit repetition codes, respectively. Such simple repetition encoding
can be useful for quantum networks as well \cite{Perseguers08b}.

Our repeater protocol can also generate high fidelity entanglement over long
distances. For example, $F^{\ast}=0.999$ and $L^{\ast}\approx730$ can be
achieved with $q=0.3\%$ and the $\left[  \left[  23,1,7\right]  \right]  $
Golay code. Such high fidelity entanglement might be useful for applications
such as quantum state teleportation and distributed quantum computation
\cite{JTSL07b}. Since quantum circuits for state-teleportation or
teleportation-based CNOT gate only use Clifford group operations, the
generated entanglement can be immediately used in these circuits without
waiting for the classical information of the Pauli frame. The adjustment of
the Pauli frame has to be postponed until the classical information is
received \footnote{Good quantum memory with coherence time longer than the
communication time might be needed.}.

Suppose good quantum memory (with coherence time longer than the communication
time) is available at the final stations, real distant Bell pairs (rather than
just strings of secret bits for quantum key distribution) can be generated.
For on-demand generation of distant Bell pairs, the time delay ($l_{0}L/c$)
associated with the classical communication to specify the Pauli frame is
inevitable, and the total time to create one Bell pair on-demand is $\tau
_{c}+l_{0}L/c$. For offline generation of distant Bell pairs that are stored
in good quantum memory for later use, we have to assume that there are enough
qubits at the final stations to store all Bell pair generated, while the
number of qubits at each intermediate station remains unchanged. Up to the
time delay ($l_{0}L/c$) for the first Bell pair , our quantum repeater channel
can create distant Bell pairs at the rate $1/\tau_{c}$, again corresponding to
the flat curve in Fig.~\ref{fig:Comparison}.

\section{Conclusion\label{sec:Conclusion}}

In summary, we have proposed a new, fast quantum repeater protocol for quantum
key distribution over intercontinental distances. Our protocol
fault-tolerantly generates a backbone of Bell pairs with CSS encoding, and
uses simple procedure of classical error correction during connection. Our
protocol using simple CSS code can provide secure quantum communication over
thousands or even millions of kilometers, with $0.3\%$ effective error
probability per physical qubit and $0.95$ target fidelity for the final Bell
pair (see Table~\ref{tab:1}). The quantum key generation rate can be above
$100$ bits/sec, only limited by the Bell pair generation between neighboring stations.

We would thank Hans Briegel, Ignacio Cirac, Wolfgang D\"{u}r, John Preskill,
Anders S\o rensen, S\'{e}bastien Perseguers, Frank Verstraete, Karl Vollbrecht
for stimulating discussions. L.J. and J.M.T. thank NII\ for the hospitality,
where part of this research was done. K.N. and W.J.M. acknowledge support in
part by MEXT, NICT, HP and QAP.

\appendix{}


\section{Effective Error Probability \label{app:EffectiveErrorProb}}

For our quantum repeater protocol, we introduce the effective error
probability $q$, which estimates the odds for obtaining a wrong output of each
physical qubit during entanglement connection. The effective error probability
combines various imperfections from entanglement generation and entanglement
connection. In the following, we will derive the effective error probability
$q$ in terms of various error parameters $\beta$, $\delta$, and $\mu$ as
detailed in Sec.~\ref{sec:IdealizedRepeater}.

First of all, we observe that all relevant operations (local CNOT gates,
teleportation-based CNOT gates, and measurements in $Z$ or $X$ basis) never
mix bit-flip errors and phase errors. For example, CNOT gates never convert
bit-flip errors into phase errors. Measurements in the $Z$ basis are only
sensitive to bit-flip errors, but not to phase errors. Therefore, we can use
two probabilities $(b,p)$ to characterize the bit-flip and phase errors,
respectively.

We will calculate these two probabilities for the physical qubits from the
operational step 1(i,ii,iii) and step 2 as illustrated in
Fig.~\ref{fig:EncodedQR}. For state distillation [step~1(i)], it is possible
to have $\left(  b^{\prime},p^{\prime}\right)  =\left(  \beta/4+\mu
/2,\beta/2+\mu/2\right)  $ for each physical qubit of the encoding block. For
entanglement purification [step~1(ii)], it is possible to have $\left(
b^{\prime\prime},p^{\prime\prime}\right)  =\left(  \beta/2,\beta/4\right)  $
for each physical qubit of the physical Bell pairs. For teleportation-based
CNOT gates \cite{Gottesman99,Zhou00,JTSL07b} [step~1(iii)], the control and
target qubits accumulate errors from the input qubits, with $\left(
b_{c}^{\prime\prime\prime},p_{c}^{\prime\prime\prime}\right)  =\left(
b^{\prime}+\beta/2,2p^{\prime}+2p^{\prime\prime}+\beta+\delta\right)  $ for
the control, and $\left(  b_{t}^{\prime\prime\prime},p_{t}^{\prime\prime
\prime}\right)  =\left(  2b^{\prime}+2b^{\prime\prime}+\beta+\delta,b^{\prime
}+\beta/2\right)  $ for the target. Finally, after entanglement connection
[step 2], the accumulated probability for obtaining a wrong output is
\begin{equation}
q_{b}=b_{c}^{\prime\prime\prime}+b_{t}^{\prime\prime\prime}+\beta
/2+\delta=\frac{15}{4}\beta+2\delta+\mu
\end{equation}
for measurements in the $Z$ basis, and is
\begin{equation}
q_{p}=p_{c}^{\prime\prime\prime}+p_{t}^{\prime\prime\prime}+\beta
/2+\delta=4\beta+2\delta+\mu
\end{equation}
for measurements in the $X$ basis. For simplicity, we may just use%
\begin{equation}
q=\max\left\{  q_{b},q_{p}\right\}  =4\beta+2\delta+\mu
\end{equation}
to estimate the effective error probability.

\section{Fault-tolerant initialization of the CSS code
\label{app:FT initialization}}

We now consider two possible approaches to fault-tolerant preparation of the
logical states $\left\vert \tilde{0}\right\rangle $ (and $\left\vert \tilde
{+}\right\rangle =\frac{1}{\sqrt{2}}\left(  \left\vert \tilde{0}\right\rangle
+\left\vert \tilde{1}\right\rangle \right)  $) of the CSS code, using local
operations within each repeater station. Both approaches use the technique of
state distillation \cite{Steane99}.

To facilitate the discussion, we first briefly review the stabilizer formulism
for the CSS code \cite{Gottesman97,NC00}. The error syndromes for the code
$C^{X}$ can be obtained by measuring the operators $\left\{  g_{j}%
^{X}\right\}  _{j=1,\cdots,m_{X}}$, and the syndromes for the code $C^{Z}$ can
be obtained by measuring the operators $\left\{  g_{j^{\prime}}^{Z}\right\}
_{j=1,\cdots,m_{Z}}$. The operators $g_{j}^{X}$ and $g_{j^{\prime}}^{Z}$
commute $\left[  g_{j}^{X},g_{j^{\prime}}^{Z}\right]  =0$ for all $j$ and
$j^{\prime}$. The operators $\left\{  g_{j}^{X}\right\}  $ and $\left\{
g_{j^{\prime}}^{Z}\right\}  $ are called the stabilizer generators. The
logical information are stored in the subspace with $+1$ eigenvalues for all
stabilizer generators $\left\{  g_{j}^{X}\right\}  $ and $\left\{
g_{j^{\prime}}^{Z}\right\}  $. (E.g., the 3-qubit repetition code is a
CSS\ code with stabilizer generators $\left\{  g_{1}^{Z},g_{2}^{Z}\right\}
=\left\{  Z_{1}Z_{2},Z_{2}Z_{3}\right\}  $; any logical state $\left\vert
\phi\right\rangle =\alpha\left\vert \tilde{0}\right\rangle +\beta\left\vert
\tilde{1}\right\rangle $ satisfies the condition $Z_{1}Z_{2}\left\vert
\phi\right\rangle =\left\vert \phi\right\rangle $ and $Z_{2}Z_{3}\left\vert
\phi\right\rangle =\left\vert \phi\right\rangle $.) Note that the stabilizer
generator $g_{j}^{Z}$ is a product of $Z$ operators, and $g_{j^{\prime}}^{X}$
is a product of $X$ operators. In addition, the logical operator $\tilde{X}$
(or $\tilde{Z}$) for the CSS code can also be expressed as a product of $X$
(or $Z$) operators. (E.g., the $3$-qubit repetition code has logical operators
$\tilde{X}=X_{1}X_{2}X_{3}$ and $\tilde{Z}=Z_{1}Z_{2}Z_{3}$.)

\subsection{First Approach}

In the first approach, we generate several copies of the logical states
$\left\vert \tilde{0}\right\rangle $, which are not fault-tolerant as the
errors might be correlated among qubits within each encoding block. For
example, one quantum gate (with error probability $\varepsilon$) may induce
errors in the multiple physical qubits; that is the probability for
multi-qubit errors can occurs at the order of $O\left(  \varepsilon\right)  $.
To suppress such multi-qubit errors, we use the state distillation circuits
(i.e., generalization of the entanglement purification circuits) to suppress
both the X and Z\ errors. After each round of distillation, the correlated
errors will be suppressed from $O\left(  \varepsilon^{l}\right)  $ to
$O\left(  \varepsilon^{l+2}+\varepsilon^{2l}\right)  $. The distillation
operation does not introduce any new correlated errors. Thus after
sufficiently many rounds of distillation, the correlated errors can be
suppressed. Meanwhile the uncorrelated errors from the distillation operations
are also suppressed by the following distillation operations. Therefore, after
sufficiently many rounds of distillation, the probability for uncorrelated
errors will reach a steady value, of the order of $\beta+\delta$ for each
physical qubit.

\subsection{Second Approach}

In the second approach, we try to avoid correlated errors from the beginning.
The idea is that we start with $n$ physical qubits initialized in the product
state $\left\vert 0\right\rangle ^{\otimes n}$, and projectively measure the
stabilizers, which can be achieved fault-tolerantly using the GHZ states (as
described in the next paragraph). We obtain a set of binary numbers associated
with the stabilizer measurements. In principle, we can perform error
correction to the encoding block to restore it to the +1 co-eigenstates for
the stabilizers. Alternatively, we may keep track of the values for the
stabilizers, and take them into account throughout the entanglement generation
and entanglement connection (as detailed below). Finally, we use several
copies of the encoding block with uncorrelated error to perform just one round
of state distillation to suppress the error probability per physical qubits to
$\sim\beta+\delta$.

To achieve fault-tolerant measurement of the stabilizer, we use $l$-qubit GHZ
states (with $l\leq n$) that can be initialized fault-tolerantly \cite{NC00}.
According to the standard form of the stabilizer code (see Ref. \cite{NC00},
page 470), the error in the value for each stabilizer is equivalent to the
error of one physical qubit.
We further improve the reliability of the stabilizer measurement by repeating
it several times \cite{Aliferis06}.

Since we have included the -1 eigenstates for the stabilizers, we need to
generalize the encoded CNOT operation by keeping track of the stabilizers as
well as the logical qubits. Suppose the encoding block for the control qubit
has eigenvalues $\left(  \mathbf{x}_{1},\mathbf{z}_{1}\right)  $ associated
with the X and Z stabilizers, and the block for the target qubit has
eigenvalues $\left(  \mathbf{x}_{2},\mathbf{z}_{2}\right)  $. The outputs have
eigenvalues $\left(  \mathbf{x}_{1},\mathbf{z}_{1}\mathbf{z}_{2}\right)  $ for
the control block and $\left(  \mathbf{x}_{1}\mathbf{x}_{2},\mathbf{z}%
_{2}\right)  $ for the target block. Consequently, when we apply the
generalized encoded CNOT\ operation to entanglement generation, there is
additional classical communication to exchange the information of stabilizers
between neighboring stations, so that both stations can update the eigenvalues
of the stabilizers for their encoding blocks. When we apply the generalized
encoded CNOT operation to entanglement connection, the classical error
correction need to take into account the eigenvalues of the stabilizers to
correct errors. Apart the these modifications, the remaining operations remain
the same.

\subsection{Estimate Local Resources for Second Approach}

We now estimate the minimum number of qubits needed for each repeater station,
which is required by the fault-tolerant preparation of the encoding block with
small error probability. (For simplicity, we assume that local operational
time is much faster than the communication time and can be safely neglected.)
We focus on the second scheme of fault-tolerant preparation, which first uses
the GHZ states to projectively measure the stabilizers and then apply state
distillation to suppress individual qubit errors. We emphasize again that both
operations of stabilizer measurement and state distillation can be performed fault-tolerantly.


The local resources are split into two categories: the memory qubits to store
two encoding blocks ($2n$ qubits), and the ancillary qubits to assist
fault-tolerant preparation. The ancillary qubits should fault-tolerantly
prepare of the GHZ state (using $n_{GHZ}$ qubits), and store additional two
encoding blocks ($2n$ qubits) for the 2-level state distillation. Altogether,
there are $4n+n_{GHZ}$ qubits for each station.

We now detail the procedure of prepare the distilled state in the storage
block $b$, using two-level state distillation with two additional blocks $a1$
and $a2$. First, we obtain a level-1 distilled encoding block in $b$ (by
projectively preparing the encoded state for $a1$ and $b$, and using $a1$ to
successfully purify $b$). Then we obtain another level-1 distilled encoding
block in $a2$ (by projectively preparing the encoded state for $a1$ and $a2$,
and using $a1$ to successfully purify $a2$). Finally, we obtain the level-2
distilled encoding block in $b$ (by using $a2$ to successfully purify $b$).
Generally, we can obtain a level-$l$ distilled block by using $l$ additional
blocks (i.e., $l~n$ qubits).

\section{Entanglement fidelity and correlation
\label{app:Fidelity and Correlation}}

There are two major sources that will reduce the entanglement fidelity for the
final encoded Bell pairs. First, the errors from the Bell measurement from
intermediate stations will lead to the wrong choice of the Pauli frame, and
the probability that all $L-2$\ Bell measurements are error-free is $\left(
1-Q\right)  ^{2\left(  L-2\right)  }$. In addition, unsuccessful local error
correction for the final encoded Bell pair will also reduce the generalized
fidelity, and the probability to have a successful error correction is
approximately $\left(  1-Q\right)  ^{2}$. Therefore, we estimate that the
entanglement fidelity to be $F\approx\left(  1-Q\right)  ^{2L-2}\gtrsim\left(
1-Q\right)  ^{2L}$.

These two sources also affect the correlation of the secret keys. If the
secret keys are obtained from the measurement in the X or Z basis, only half
of the $2\left(  L-2\right)  $\ classical bits from intermediate repeater
stations are relevant while the other half do not affect the keys at all. And
the probability for successful classical error correction to infer the encoded
logical qubit is of the order of $\left(  1-Q\right)  ^{2}$. Therefore, the
correlation is approximately $C\approx\left(  1-Q\right)  ^{L}\approx\sqrt{F}$.

\section{Time Overhead and Failure Probability for Entanglement
Purification\label{app:TimeOverhead}}

We now consider the process of generating $n$ purified Bell pairs between
neighboring stations. We will calculate the failure probability $P_{fail}$ for
obtaining at least $n$ purified Bell pairs using $N_{0}$ unpurified Bell
pairs. The failure probability should also depend on the fidelity of
unpurified Bell pairs ($F_{0}$) and the error probability for local operations
($\beta$ and $\delta$). Generally, the more unpurified Bell pairs $N_{0}$, and
the smaller failure probability $P_{fail}$. For a given $P_{fail}$, we can
estimate the $N_{0}$ and consequently the cycle time $\tau_{c}$ that
determines the key generation rate.

\subsection{Failure Probability}

\begin{figure}[tb]
\begin{center}
\includegraphics[
width=6cm
]{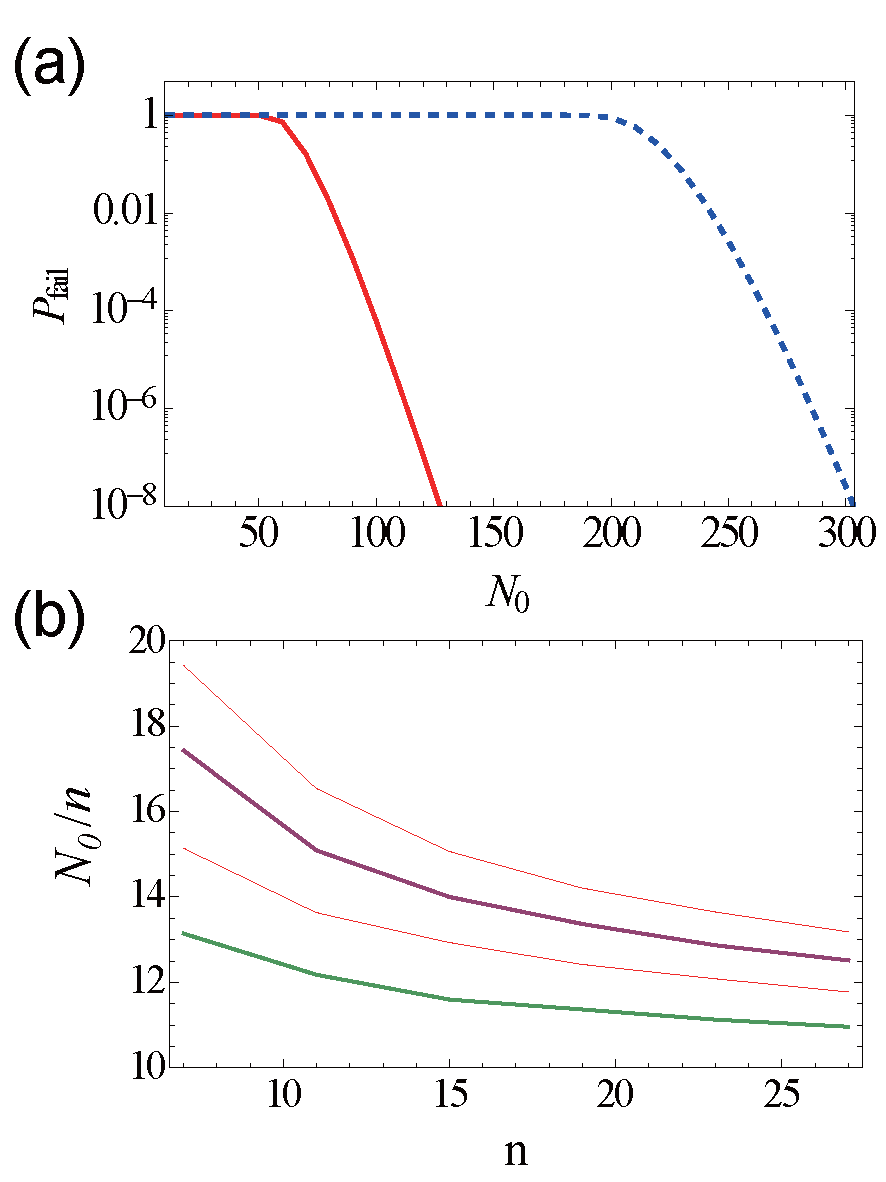}
\end{center}
\caption[fig:FailProb]{Failure probability and unpurified Bell pairs. (a) The
failure probability $P_{fail}$ decreases exponentially with the number of
unpurified Bell pairs $N_{0}$ (when $N_{0}$ surpasses certain threshold), for
$n=7$ (red solid line) and $n=23$ (blue dashed line). (b) For fixed $P_{fail}%
$, the ratio $N_{0}/n\sim15$ for a wide range of $n$. The four curves from
lower left to the upper right correspond to $P_{fail}=10^{-3},10^{-5},10^{-7}$
and $10^{-9}$, respectively. For both plots, we assume unpurified Bell pairs
with fidelity $F_{0}=0.95$ due to depolarizing error. The operational error
probabilities are $\beta=\delta=10^{-3}$. After three levels of purifications,
the fidelity of the Bell pair can be $0.9984$. }%
\label{fig:FailProb}%
\end{figure}

In order to obtain the failure probability, we first calculate the number
distribution for purified Bell pairs obtained from $N_{0}$ unpurified Bell pairs.

We distinguish the purified Bell pairs according their level of purification.
A level-$\left(  i+1\right)  $ pair is obtained from a successful purification
using two level-$i$ pairs. Level-$0$ pairs are the same as unpurified Bell
pairs. Level-$l$ pairs are directly used for non-local CNOT\ gates.

We introduce the number distribution $\left\{  p_{m}^{\left(  i\right)
}\right\}  _{m=0,1,2,\cdots}$ for level-$i$ pairs obtained from $N_{0}$
unpurified Bell pairs, with $i=0,1,\cdots,l$. The number distribution for
level-$0$ pairs is%
\begin{equation}
p_{m}^{\left(  0\right)  }=\delta_{m,N_{0}}.
\end{equation}
As two level-$i$ pairs are needed for one level-$\left(  i+1\right)  $ pair,
we define%
\begin{equation}
\tilde{p}_{k}^{\left(  i\right)  }=p_{2k}^{\left(  i\right)  }+p_{2k+1}%
^{\left(  i\right)  },
\end{equation}
which can be used to calculate the number distribution for level-$\left(
i+1\right)  $ pairs%
\begin{equation}
p_{m}^{\left(  i+1\right)  }=\sum_{j=m}\left(
\begin{array}
[c]{c}%
j\\
m
\end{array}
\right)  r_{i}^{m}\left(  1-r_{i}\right)  ^{j-m}\tilde{p}_{j}^{\left(
i\right)  }\text{,}%
\end{equation}
where $r_{i}$ is the success probability for obtaining a level-$\left(
i+1\right)  $ pair from two level-$i$ pairs. Thus, the failure probability is%
\begin{equation}
P_{fail}=\sum_{j=0}^{n-1}p_{j}^{\left(  l\right)  }. \label{eq:Pfail}%
\end{equation}

For example, given $\beta=\delta=10^{-3}$ and $F_{0}=0.95$ with depolarizing
error, the fidelity for level-$3$ purified pair can be $0.9984$. In
Fig.~\ref{fig:FailProb}(a), we plot the failure probability that decreases
exponentially when $N_{0}$ surpasses certain threshold. In
Fig.~\ref{fig:FailProb}(b), we plot $N_{0}/n$ as a function of $n$, requiring
fixed failure probability $P_{fail}$ ($10^{-3},10^{-5}$, $10^{-7}$ or
$10^{-9}$). We note that $N_{0}/n\approx15$ is sufficient to ensure
$P_{fail}<10^{-5}$ a wide range of $n$.

\subsection{Time Overhead and Key Generate Rate}

We now estimate the time needed to obtain $n$ purified Bell pairs between two
neighboring repeater stations. Each attempt of entanglement generation takes
time $l_{0}/v$, with success probability $\eta^{2}e^{-l_{0}/l_{att}}$. Since
there are $n_{EnG}$ ($=2n+n_{GHZ}$) qubits available at each station, the
generation rate of unpurified Bell pairs is%
\begin{equation}
R=\frac{v}{l_{0}}\eta^{2}e^{-l_{0}/l_{att}}n_{EnG},
\end{equation}
where the spacing between nearest stations is $l_{0}=10$ km, the fiber
attenuation length is $l_{att}=20$ km, the signal propagation speed is
$v=2\times10^{5}$ km/s, and the overall efficiency for collecting and
detecting single photon is $\eta\approx0.3$. We have $R=n_{EnG}1.1\times
10^{3}\sec^{-1}$.

We can estimate the time to obtain $N_{0}$ unpurified Bell pairs $\tau
_{0}=N_{0}/R$. Since each station need to connect with both neighboring
stations, the total cycle time is twice as long:%
\begin{equation}
\tau_{c}=2N_{0}/R=\kappa\frac{l_{0}}{v}\frac{e^{l_{0}/l_{att}}}{\eta^{2}},
\end{equation}
with%
\begin{equation}
\kappa=\frac{2N_{0}}{n_{EnG}}\approx\frac{2N_{0}}{4n}\approx8,
\end{equation}
where the last equality assumes $n_{EnG}\approx4n$ (i.e., $n_{GHZ}\approx2n$)
and $N_{0}/n\approx15$ to ensure $P_{fail}<10^{-5}$ [see
Fig.~\ref{fig:FailProb}(b)]. Therefore, for the parameters considered here,
approximately $6n$ qubits at each station can achieve $\tau_{c}\approx7$ ms,
which is sufficient for quantum key generation rate of $100$ bits/sec over
long distances.

\bibliographystyle{apsrev}
\bibliography{ref}

\begin{thebibliography}{31}
\expandafter\ifx\csname natexlab\endcsname\relax\def\natexlab#1{#1}\fi
\expandafter\ifx\csname bibnamefont\endcsname\relax
  \def\bibnamefont#1{#1}\fi
\expandafter\ifx\csname bibfnamefont\endcsname\relax
  \def\bibfnamefont#1{#1}\fi
\expandafter\ifx\csname citenamefont\endcsname\relax
  \def\citenamefont#1{#1}\fi
\expandafter\ifx\csname url\endcsname\relax
  \def\url#1{\texttt{#1}}\fi
\expandafter\ifx\csname urlprefix\endcsname\relax\def\urlprefix{URL }\fi
\providecommand{\bibinfo}[2]{#2}
\providecommand{\eprint}[2][]{\url{#2}}

\bibitem[{\citenamefont{Gisin et~al.}(2002)\citenamefont{Gisin, Ribordy,
  Tittel, and Zbinden}}]{Gisin02}
\bibinfo{author}{\bibfnamefont{N.}~\bibnamefont{Gisin}},
  \bibinfo{author}{\bibfnamefont{G.~G.} \bibnamefont{Ribordy}},
  \bibinfo{author}{\bibfnamefont{W.}~\bibnamefont{Tittel}}, \bibnamefont{and}
  \bibinfo{author}{\bibfnamefont{H.}~\bibnamefont{Zbinden}},
  \bibinfo{journal}{Rev. Mod. Phys.} \textbf{\bibinfo{volume}{74}},
  \bibinfo{pages}{145} (\bibinfo{year}{2002}).

\bibitem[{\citenamefont{Ursin et~al.}(2007)\citenamefont{Ursin, Tiefenbacher,
  Schmitt-Manderbach, Weier, Scheidl, Lindenthal, Blauensteiner, Jennewein,
  Perdigues, Trojek et~al.}}]{Ursin07}
\bibinfo{author}{\bibfnamefont{R.}~\bibnamefont{Ursin}},
  \bibinfo{author}{\bibfnamefont{F.}~\bibnamefont{Tiefenbacher}},
  \bibinfo{author}{\bibfnamefont{T.}~\bibnamefont{Schmitt-Manderbach}},
  \bibinfo{author}{\bibfnamefont{H.}~\bibnamefont{Weier}},
  \bibinfo{author}{\bibfnamefont{T.}~\bibnamefont{Scheidl}},
  \bibinfo{author}{\bibfnamefont{M.}~\bibnamefont{Lindenthal}},
  \bibinfo{author}{\bibfnamefont{B.}~\bibnamefont{Blauensteiner}},
  \bibinfo{author}{\bibfnamefont{T.}~\bibnamefont{Jennewein}},
  \bibinfo{author}{\bibfnamefont{J.}~\bibnamefont{Perdigues}},
  \bibinfo{author}{\bibfnamefont{P.}~\bibnamefont{Trojek}},
  \bibnamefont{et~al.}, \bibinfo{journal}{Nature Phys.}
  \textbf{\bibinfo{volume}{3}}, \bibinfo{pages}{481} (\bibinfo{year}{2007}).

\bibitem[{\citenamefont{Briegel et~al.}(1998)\citenamefont{Briegel, Dur, Cirac,
  and Zoller}}]{Briegel98}
\bibinfo{author}{\bibfnamefont{H.~J.} \bibnamefont{Briegel}},
  \bibinfo{author}{\bibfnamefont{W.}~\bibnamefont{Dur}},
  \bibinfo{author}{\bibfnamefont{J.~I.} \bibnamefont{Cirac}}, \bibnamefont{and}
  \bibinfo{author}{\bibfnamefont{P.}~\bibnamefont{Zoller}},
  \bibinfo{journal}{Phys. Rev. Lett.} \textbf{\bibinfo{volume}{81}},
  \bibinfo{pages}{5932} (\bibinfo{year}{1998}).

\bibitem[{\citenamefont{Childress et~al.}(2006)\citenamefont{Childress, Taylor,
  Sorensen, and Lukin}}]{Childress06b}
\bibinfo{author}{\bibfnamefont{L.}~\bibnamefont{Childress}},
  \bibinfo{author}{\bibfnamefont{J.~M.} \bibnamefont{Taylor}},
  \bibinfo{author}{\bibfnamefont{A.~S.} \bibnamefont{Sorensen}},
  \bibnamefont{and} \bibinfo{author}{\bibfnamefont{M.~D.} \bibnamefont{Lukin}},
  \bibinfo{journal}{Phys. Rev. Lett.} \textbf{\bibinfo{volume}{96}},
  \bibinfo{pages}{070504} (\bibinfo{year}{2006}).

\bibitem[{\citenamefont{van Loock et~al.}(2006)\citenamefont{van Loock, Ladd,
  Sanaka, Yamaguchi, Nemoto, Munro, and Yamamoto}}]{vanLoock06}
\bibinfo{author}{\bibfnamefont{P.}~\bibnamefont{van Loock}},
  \bibinfo{author}{\bibfnamefont{T.~D.} \bibnamefont{Ladd}},
  \bibinfo{author}{\bibfnamefont{K.}~\bibnamefont{Sanaka}},
  \bibinfo{author}{\bibfnamefont{F.}~\bibnamefont{Yamaguchi}},
  \bibinfo{author}{\bibfnamefont{K.}~\bibnamefont{Nemoto}},
  \bibinfo{author}{\bibfnamefont{W.~J.} \bibnamefont{Munro}}, \bibnamefont{and}
  \bibinfo{author}{\bibfnamefont{Y.}~\bibnamefont{Yamamoto}},
  \bibinfo{journal}{Phys. Rev. Lett.} \textbf{\bibinfo{volume}{96}},
  \bibinfo{pages}{240501} (\bibinfo{year}{2006}).

\bibitem[{\citenamefont{Jiang et~al.}(2007{\natexlab{a}})\citenamefont{Jiang,
  Taylor, Khaneja, and Lukin}}]{JTKL07}
\bibinfo{author}{\bibfnamefont{L.}~\bibnamefont{Jiang}},
  \bibinfo{author}{\bibfnamefont{J.~M.} \bibnamefont{Taylor}},
  \bibinfo{author}{\bibfnamefont{N.}~\bibnamefont{Khaneja}}, \bibnamefont{and}
  \bibinfo{author}{\bibfnamefont{M.~D.} \bibnamefont{Lukin}},
  \bibinfo{journal}{Proc. Natl. Acad. Sci. U. S. A.}
  \textbf{\bibinfo{volume}{104}}, \bibinfo{pages}{17291}
  (\bibinfo{year}{2007}{\natexlab{a}}).

\bibitem[{\citenamefont{Collins et~al.}(2007)\citenamefont{Collins, Jenkins,
  Kuzmich, and Kennedy}}]{Collins07}
\bibinfo{author}{\bibfnamefont{O.~A.} \bibnamefont{Collins}},
  \bibinfo{author}{\bibfnamefont{S.~D.} \bibnamefont{Jenkins}},
  \bibinfo{author}{\bibfnamefont{A.}~\bibnamefont{Kuzmich}}, \bibnamefont{and}
  \bibinfo{author}{\bibfnamefont{T.~A.~B.} \bibnamefont{Kennedy}},
  \bibinfo{journal}{Phys. Rev. Lett.} \textbf{\bibinfo{volume}{98}},
  \bibinfo{pages}{060502} (\bibinfo{year}{2007}).

\bibitem[{\citenamefont{Van~Meter et~al.}(2007)\citenamefont{Van~Meter, Ladd,
  Munro, and Nemoto}}]{VanMeter07c}
\bibinfo{author}{\bibfnamefont{R.}~\bibnamefont{Van~Meter}},
  \bibinfo{author}{\bibfnamefont{T.~D.} \bibnamefont{Ladd}},
  \bibinfo{author}{\bibfnamefont{W.~J.} \bibnamefont{Munro}}, \bibnamefont{and}
  \bibinfo{author}{\bibfnamefont{K.}~\bibnamefont{Nemoto}},
  \bibinfo{journal}{e-print arXiv:} \textbf{\bibinfo{volume}{0705.4128}}
  (\bibinfo{year}{2007}).

\bibitem[{\citenamefont{Hartmann et~al.}(2007)\citenamefont{Hartmann, Kraus,
  Briegel, and Dur}}]{HartmannL07}
\bibinfo{author}{\bibfnamefont{L.}~\bibnamefont{Hartmann}},
  \bibinfo{author}{\bibfnamefont{B.}~\bibnamefont{Kraus}},
  \bibinfo{author}{\bibfnamefont{H.~J.} \bibnamefont{Briegel}},
  \bibnamefont{and} \bibinfo{author}{\bibfnamefont{W.}~\bibnamefont{Dur}},
  \bibinfo{journal}{Phys. Rev. A} \textbf{\bibinfo{volume}{75}},
  \bibinfo{pages}{032310} (\bibinfo{year}{2007}).

\bibitem[{\citenamefont{Nielsen and Chuang}(2000)}]{NC00}
\bibinfo{author}{\bibfnamefont{M.~A.} \bibnamefont{Nielsen}} \bibnamefont{and}
  \bibinfo{author}{\bibfnamefont{I.}~\bibnamefont{Chuang}},
  \emph{\bibinfo{title}{Quantum computation and quantum information}}
  (\bibinfo{publisher}{Cambridge University Press},
  \bibinfo{address}{Cambridge, U.K; New York}, \bibinfo{year}{2000}).

\bibitem[{\citenamefont{Dur et~al.}(1999)\citenamefont{Dur, Briegel, Cirac, and
  Zoller}}]{Dur99}
\bibinfo{author}{\bibfnamefont{W.}~\bibnamefont{Dur}},
  \bibinfo{author}{\bibfnamefont{H.~J.} \bibnamefont{Briegel}},
  \bibinfo{author}{\bibfnamefont{J.~I.} \bibnamefont{Cirac}}, \bibnamefont{and}
  \bibinfo{author}{\bibfnamefont{P.}~\bibnamefont{Zoller}},
  \bibinfo{journal}{Phys. Rev. A} \textbf{\bibinfo{volume}{59}},
  \bibinfo{pages}{169} (\bibinfo{year}{1999}).

\bibitem[{\citenamefont{Bennett et~al.}(1993)\citenamefont{Bennett, Brassard,
  Crepeau, Jozsa, Peres, and Wootters}}]{Bennett93}
\bibinfo{author}{\bibfnamefont{C.~H.} \bibnamefont{Bennett}},
  \bibinfo{author}{\bibfnamefont{G.}~\bibnamefont{Brassard}},
  \bibinfo{author}{\bibfnamefont{C.}~\bibnamefont{Crepeau}},
  \bibinfo{author}{\bibfnamefont{R.}~\bibnamefont{Jozsa}},
  \bibinfo{author}{\bibfnamefont{A.}~\bibnamefont{Peres}}, \bibnamefont{and}
  \bibinfo{author}{\bibfnamefont{W.~K.} \bibnamefont{Wootters}},
  \bibinfo{journal}{Phys. Rev. Lett.} \textbf{\bibinfo{volume}{70}},
  \bibinfo{pages}{1895} (\bibinfo{year}{1993}).

\bibitem[{\citenamefont{Zukowski et~al.}(1993)\citenamefont{Zukowski,
  Zeilinger, Horne, and Ekert}}]{Zukowski93}
\bibinfo{author}{\bibfnamefont{M.}~\bibnamefont{Zukowski}},
  \bibinfo{author}{\bibfnamefont{A.}~\bibnamefont{Zeilinger}},
  \bibinfo{author}{\bibfnamefont{M.~A.} \bibnamefont{Horne}}, \bibnamefont{and}
  \bibinfo{author}{\bibfnamefont{A.~K.} \bibnamefont{Ekert}},
  \bibinfo{journal}{Phys. Rev. Lett.} \textbf{\bibinfo{volume}{71}},
  \bibinfo{pages}{4287} (\bibinfo{year}{1993}).

\bibitem[{\citenamefont{Knill}(2005)}]{Knill05}
\bibinfo{author}{\bibfnamefont{E.}~\bibnamefont{Knill}},
  \bibinfo{journal}{Nature (London)} \textbf{\bibinfo{volume}{434}},
  \bibinfo{pages}{39} (\bibinfo{year}{2005}).

\bibitem[{\citenamefont{Ekert}(1991)}]{Ekert91}
\bibinfo{author}{\bibfnamefont{A.~K.} \bibnamefont{Ekert}},
  \bibinfo{journal}{Phys. Rev. Lett.} \textbf{\bibinfo{volume}{67}},
  \bibinfo{pages}{661} (\bibinfo{year}{1991}).

\bibitem[{\citenamefont{Gottesman and Chuang}(1999)}]{Gottesman99}
\bibinfo{author}{\bibfnamefont{D.}~\bibnamefont{Gottesman}} \bibnamefont{and}
  \bibinfo{author}{\bibfnamefont{I.~L.} \bibnamefont{Chuang}},
  \bibinfo{journal}{Nature (London)} \textbf{\bibinfo{volume}{402}},
  \bibinfo{pages}{390} (\bibinfo{year}{1999}).

\bibitem[{\citenamefont{Zhou et~al.}(2000)\citenamefont{Zhou, Leung, and
  Chuang}}]{Zhou00}
\bibinfo{author}{\bibfnamefont{X.}~\bibnamefont{Zhou}},
  \bibinfo{author}{\bibfnamefont{D.~W.} \bibnamefont{Leung}}, \bibnamefont{and}
  \bibinfo{author}{\bibfnamefont{I.~L.} \bibnamefont{Chuang}},
  \bibinfo{journal}{Phys. Rev. A} \textbf{\bibinfo{volume}{62}},
  \bibinfo{pages}{052316} (\bibinfo{year}{2000}).

\bibitem[{\citenamefont{Jiang et~al.}(2007{\natexlab{b}})\citenamefont{Jiang,
  Taylor, Sorensen, and Lukin}}]{JTSL07b}
\bibinfo{author}{\bibfnamefont{L.}~\bibnamefont{Jiang}},
  \bibinfo{author}{\bibfnamefont{J.~M.} \bibnamefont{Taylor}},
  \bibinfo{author}{\bibfnamefont{A.~S.} \bibnamefont{Sorensen}},
  \bibnamefont{and} \bibinfo{author}{\bibfnamefont{M.~D.} \bibnamefont{Lukin}},
  \bibinfo{journal}{Phys. Rev. A} \textbf{\bibinfo{volume}{76}},
  \bibinfo{pages}{062323} (\bibinfo{year}{2007}{\natexlab{b}}).

\bibitem[{\citenamefont{Shor and Preskill}(2000)}]{Shor00}
\bibinfo{author}{\bibfnamefont{P.~W.} \bibnamefont{Shor}} \bibnamefont{and}
  \bibinfo{author}{\bibfnamefont{J.}~\bibnamefont{Preskill}},
  \bibinfo{journal}{Phys. Rev. Lett.} \textbf{\bibinfo{volume}{85}},
  \bibinfo{pages}{441} (\bibinfo{year}{2000}).

\bibitem[{\citenamefont{Steane}(2003)}]{Steane03}
\bibinfo{author}{\bibfnamefont{A.~M.} \bibnamefont{Steane}},
  \bibinfo{journal}{Phys. Rev. A} \textbf{\bibinfo{volume}{68}},
  \bibinfo{pages}{042322} (\bibinfo{year}{2003}).

\bibitem[{\citenamefont{Calderbank and Shor}(1996)}]{Canderbank96}
\bibinfo{author}{\bibfnamefont{A.~R.} \bibnamefont{Calderbank}}
  \bibnamefont{and} \bibinfo{author}{\bibfnamefont{P.~W.} \bibnamefont{Shor}},
  \bibinfo{journal}{Phys. Rev. A} \textbf{\bibinfo{volume}{54}},
  \bibinfo{pages}{1098} (\bibinfo{year}{1996}).

\bibitem[{\citenamefont{Leibfried et~al.}(2004)\citenamefont{Leibfried,
  Barrett, Schaetz, Britton, Chiaverini, Itano, Jost, Langer, and
  Wineland}}]{Leibfried04}
\bibinfo{author}{\bibfnamefont{D.}~\bibnamefont{Leibfried}},
  \bibinfo{author}{\bibfnamefont{M.~D.} \bibnamefont{Barrett}},
  \bibinfo{author}{\bibfnamefont{T.}~\bibnamefont{Schaetz}},
  \bibinfo{author}{\bibfnamefont{J.}~\bibnamefont{Britton}},
  \bibinfo{author}{\bibfnamefont{J.}~\bibnamefont{Chiaverini}},
  \bibinfo{author}{\bibfnamefont{W.~M.} \bibnamefont{Itano}},
  \bibinfo{author}{\bibfnamefont{J.~D.} \bibnamefont{Jost}},
  \bibinfo{author}{\bibfnamefont{C.}~\bibnamefont{Langer}}, \bibnamefont{and}
  \bibinfo{author}{\bibfnamefont{D.~J.} \bibnamefont{Wineland}},
  \bibinfo{journal}{Science} \textbf{\bibinfo{volume}{304}},
  \bibinfo{pages}{1476} (\bibinfo{year}{2004}).

\bibitem[{\citenamefont{Riebe et~al.}(2004)\citenamefont{Riebe, Haffner, Roos,
  Hansel, Benhelm, Lancaster, Korber, Becher, Schmidt-Kaler, James
  et~al.}}]{Riebe04}
\bibinfo{author}{\bibfnamefont{M.}~\bibnamefont{Riebe}},
  \bibinfo{author}{\bibfnamefont{H.}~\bibnamefont{Haffner}},
  \bibinfo{author}{\bibfnamefont{C.~F.} \bibnamefont{Roos}},
  \bibinfo{author}{\bibfnamefont{W.}~\bibnamefont{Hansel}},
  \bibinfo{author}{\bibfnamefont{J.}~\bibnamefont{Benhelm}},
  \bibinfo{author}{\bibfnamefont{G.~P.~T.} \bibnamefont{Lancaster}},
  \bibinfo{author}{\bibfnamefont{T.~W.} \bibnamefont{Korber}},
  \bibinfo{author}{\bibfnamefont{C.}~\bibnamefont{Becher}},
  \bibinfo{author}{\bibfnamefont{F.}~\bibnamefont{Schmidt-Kaler}},
  \bibinfo{author}{\bibfnamefont{D.~F.~V.} \bibnamefont{James}},
  \bibnamefont{et~al.}, \bibinfo{journal}{Nature (London)}
  \textbf{\bibinfo{volume}{429}}, \bibinfo{pages}{734} (\bibinfo{year}{2004}).

\bibitem[{\citenamefont{Jelezko et~al.}(2004)\citenamefont{Jelezko, Gaebel,
  Popa, Domhan, Gruber, and Wrachtrup}}]{Jelezko04}
\bibinfo{author}{\bibfnamefont{F.}~\bibnamefont{Jelezko}},
  \bibinfo{author}{\bibfnamefont{T.}~\bibnamefont{Gaebel}},
  \bibinfo{author}{\bibfnamefont{I.}~\bibnamefont{Popa}},
  \bibinfo{author}{\bibfnamefont{M.}~\bibnamefont{Domhan}},
  \bibinfo{author}{\bibfnamefont{A.}~\bibnamefont{Gruber}}, \bibnamefont{and}
  \bibinfo{author}{\bibfnamefont{J.}~\bibnamefont{Wrachtrup}},
  \bibinfo{journal}{Phys. Rev. Lett.} \textbf{\bibinfo{volume}{93}},
  \bibinfo{pages}{130501} (\bibinfo{year}{2004}).

\bibitem[{\citenamefont{Dutt et~al.}(2007)\citenamefont{Dutt, Childress, Jiang,
  Togan, Maze, Jelezko, Zibrov, Hemmer, and Lukin}}]{Dutt07}
\bibinfo{author}{\bibfnamefont{M.~V.~G.} \bibnamefont{Dutt}},
  \bibinfo{author}{\bibfnamefont{L.}~\bibnamefont{Childress}},
  \bibinfo{author}{\bibfnamefont{L.}~\bibnamefont{Jiang}},
  \bibinfo{author}{\bibfnamefont{E.}~\bibnamefont{Togan}},
  \bibinfo{author}{\bibfnamefont{J.}~\bibnamefont{Maze}},
  \bibinfo{author}{\bibfnamefont{F.}~\bibnamefont{Jelezko}},
  \bibinfo{author}{\bibfnamefont{A.~S.} \bibnamefont{Zibrov}},
  \bibinfo{author}{\bibfnamefont{P.~R.} \bibnamefont{Hemmer}},
  \bibnamefont{and} \bibinfo{author}{\bibfnamefont{M.~D.} \bibnamefont{Lukin}},
  \bibinfo{journal}{Science} \textbf{\bibinfo{volume}{316}},
  \bibinfo{pages}{1312} (\bibinfo{year}{2007}).

\bibitem[{\citenamefont{Bacon}(2006)}]{Bacon06}
\bibinfo{author}{\bibfnamefont{D.}~\bibnamefont{Bacon}},
  \bibinfo{journal}{Phys. Rev. A} \textbf{\bibinfo{volume}{73}},
  \bibinfo{pages}{012340} (\bibinfo{year}{2006}).

\bibitem[{\citenamefont{Perseguers et~al.}(2008)\citenamefont{Perseguers,
  Jiang, Schuch, Verstraete, Lukin, Cirac, and Vollbrecht}}]{Perseguers08b}
\bibinfo{author}{\bibfnamefont{S.}~\bibnamefont{Perseguers}},
  \bibinfo{author}{\bibfnamefont{L.}~\bibnamefont{Jiang}},
  \bibinfo{author}{\bibfnamefont{N.}~\bibnamefont{Schuch}},
  \bibinfo{author}{\bibfnamefont{F.}~\bibnamefont{Verstraete}},
  \bibinfo{author}{\bibfnamefont{M.~D.} \bibnamefont{Lukin}},
  \bibinfo{author}{\bibfnamefont{J.~I.} \bibnamefont{Cirac}}, \bibnamefont{and}
  \bibinfo{author}{\bibfnamefont{K.~G.~H.} \bibnamefont{Vollbrecht}},
  \bibinfo{journal}{Phys. Rev. A} \textbf{\bibinfo{volume}{78}},
  \bibinfo{pages}{062324} (\bibinfo{year}{2008}).

\bibitem[{\citenamefont{Steane}(1999)}]{Steane99}
\bibinfo{author}{\bibfnamefont{A.~M.} \bibnamefont{Steane}},
  \bibinfo{journal}{Nature (London)} \textbf{\bibinfo{volume}{399}},
  \bibinfo{pages}{124} (\bibinfo{year}{1999}).

\bibitem[{\citenamefont{Gottesman}(1997)}]{Gottesman97}
\bibinfo{author}{\bibfnamefont{D.}~\bibnamefont{Gottesman}}, Ph.D. thesis,
  \bibinfo{school}{Caltech} (\bibinfo{year}{1997}).

\bibitem[{\citenamefont{Aliferis et~al.}(2006)\citenamefont{Aliferis,
  Gottesman, and Preskill}}]{Aliferis06}
\bibinfo{author}{\bibfnamefont{P.}~\bibnamefont{Aliferis}},
  \bibinfo{author}{\bibfnamefont{D.}~\bibnamefont{Gottesman}},
  \bibnamefont{and} \bibinfo{author}{\bibfnamefont{J.}~\bibnamefont{Preskill}},
  \bibinfo{journal}{Quantum Inf. Comput.} \textbf{\bibinfo{volume}{6}},
  \bibinfo{pages}{97} (\bibinfo{year}{2006}).

\bibitem[{\citenamefont{Lo and Chau}(1999)}]{Lo99}
\bibinfo{author}{\bibfnamefont{H.-K.} \bibnamefont{Lo}} \bibnamefont{and}
  \bibinfo{author}{\bibfnamefont{H.~F.} \bibnamefont{Chau}},
  \bibinfo{journal}{Science} \textbf{\bibinfo{volume}{283}},
  \bibinfo{pages}{2050} (\bibinfo{year}{1999}).

\end{thebibliography}

\end{document}